\documentclass[aps,prd,groupedaddress,showpacs,showkeys]{revtex4}
\usepackage{graphics}

\begin{document}
\title{Lifetime of Kaonium}

\author{S. Krewald}
\author{R. H. Lemmer\footnotemark[1]\footnotetext[1]{Permanent address:
Nuclear and Particle Theory Group, University of the Witwatersrand,
Johannesburg, Private Bag 3, WITS 2050, South Africa.}}
\author{F. P. Sassen}

\affiliation{Institut f\"ur Kernphysik,\\
 Forschungszentrum J{\"u}lich GmbH, D--52425,
J{\"u}lich, Germany.}

\date{\today}

\begin{abstract}
 The kaon--antikaon system is studied in both the atomic and the 
strongly interacting sector. We discuss the influence of the structures
of the $f_0(980)$ and the $a_0(980)$ mesons on the lifetime of
kaonium. The strong interactions are generated by
  vector meson exchange within the framework of the standard 
  $SU(3)_V\times SU(3)_A$ invariant effective Lagrangian. 
In the atomic sector, the
energy levels and decay widths of kaonium are determined by an eigenvalue 
equation of the Kudryavtsev--Popov  type, with the strong interaction effects 
 entering through
the complex scattering length for $K\bar K$ scattering and annihilation.   
The presence of two scalar mesons, $f_0(980)$ and
$a_0(980)$, leads to a ground state energy for the kaonium atom that is shifted above  the
    point Coulomb value by a few hundred eV. The effect   
     on the lifetime for the kaonium decay into two pions
    is much more dramatic. This lifetime is reduced by two orders of magnitude from
    $1.2\times 10^{-16}$ sec for annihilation in  a pure Coulomb field down 
    to $3.2\times 10^{-18}$ sec when the strong interactions 
    are included. 
The analysis of the two photon decay width of the $f_0(980)$
suggests a generalization of the molecular picture which reduces the
lifetime of kaonium still further to $1.1\times 10^{-18}\textrm{ sec}$.

\end{abstract}


\maketitle
\section{Introduction}

There has been  substantial experimental progress in the field of
meson spectroscopy during the last decade
\cite{amsler}.
In the energy region up to 2 GeV, more scalar-isoscalar mesons
have been established than can be accounted for by a
quark-antiquark structure
\cite{godfrey}.
The structure of the scalar meson with lowest mass, the $f_0(980)$, has
been controversial for many years. The $f_0(980)$ might be a $q\bar{q}$ state
\cite{Montanet2000,Anisovich2000},
a $q^2 \bar{q}^2$ state
\cite{achasov82},
 or a $K\bar{K}$ molecule
\cite{WS90}.

The radiative decay of the $\phi$-meson provides a particularly strong argument
for a  $q^2 \bar{q}^2$ interpretation of both the $f_0(980)$ and the $a_0(980)$ 
mesons, as was first pointed out by N.N. Achasov and V.N. Ivanchenko\cite{achasov82}.
Both the recent Novosibirsk data\cite{achassov00} and the KLOE data\cite{KLOE} can
be reproduced by a calculation which generates those mesons dynamically, however\cite{oller03}.
Since in Ref.\cite{oller03}
 Oller had to introduce a $\phi \rightarrow \gamma K^0 \bar{K}^0$ contact interaction, 
 the issue remains controversial\cite{achasov03}.

The  production of two neutral pions in ultrarelativistic pion proton reactions
shows a strong dependence of the S-wave amplitude on the momentum transferred between
the proton and the neutron for invariant two-pion masses in the vicinity of 1 GeV\cite{gams,gunter01}.
This fact has been interpreted as evidence for a hard component of the $f_0(980)$,
see e.g. Ref. \cite{Achasov1}. A recent calculation based on a model which allows for a dynamical
generation of the $ f_0(980)$ provides a good description of the data\cite{sassen03}. On the
other hand, the model employed in Ref.\cite{sassen03} also includes a bare scalar resonance.
Given this situation, we feel that a simplified calculation might be helpful.

In this paper, we develop an analytical model which generates the $f_0(980)$ meson
as a bound $K\bar{K}$ structure. This model makes specific predictions for the
structure of the exotic atom Kaonium.  In a second step, we work out the
predictions for Kaonium based on the meson-exchange model of Ref.\cite{sassen03}.

 The molecular interpretation  is consistent with the
 small  binding energy $\sim 10$--$20$ MeV of the $K\bar K$ system relative
 to its reduced mass $M_K\approx 496$ MeV. This suggests a non--relativisic
 effective field theory approach\cite{LLR02}  that has  also been used
 recently            
 to study both  pionium\cite{GLRG01,HadAtom02} as well as ${\cal O}(\alpha^6)$ QED
 recoil and radiative corrections to the positronium spectrum\cite{CMY99}.
 With this in mind we  use the standard $SU(3)_V\times SU(3)_A$
 Lagrangian\cite{deSwart63} to describe the dynamics of the
 $K\bar K$ interaction\cite{Lohse90,Jansen95} and decay via the exchange of
 $\rho,\omega,\phi,K^*....$ vector mesons.  The coupling constants appearing in the Lagrangian
 are related via  $SU(3)$ symmetry to the
 $\rho\pi\pi$ coupling constant
 $g_{\rho\pi\pi}$, which in turn can be  obtained
  from the KSRF relation\cite{KSRF66}. Thus once the decision on the
  form of Lagrangian has been taken, only physical meson masses and known coupling constants
  enter into the calculations.

 We work in the non--relativistic limit in which case two
 important simplifications  occur: (i) only $t$--channel exchange diagrams
 survive,
 and
 (ii) the resulting one--meson exchange
 potentials become local.  This  in turn means that one can reduce the
 Bethe--Salpeter (BS)
 integral equation for the bound states of the interacting $K\bar K$ system to a local two--body
 Schroedinger wave equation that offers a significant
  simplification over working with integral equations\cite{Lohse90,Sassen01}.

 After a brief recall of the derivation of
 the wave equation
 for non--relativistic local potentials
 from the BS equation in Section II, the calculation of
 one meson exchange  potentials involving  both
 direct $\rho,\omega,\phi$ transfer between $K$ and $\bar K$  as well as
  $K\bar K\to K\bar K$ scattering via two--pion intermediate states
  involving $K^*$ exchanges is carried out in Section III.
  We refer to these potentials collectively
  as one--boson exchange (OBE) potentials.
 The last contribution is essential to describe the
  $K\bar K\to  2\pi$  decay channel.
  Then we make use of the fact that one can describe the
 low--energy
 properties of $K\bar K$ very adequately in the effective range
 approximation, to replace the
 OBE potentials by phase--equivalent potentials of the Bargmann
 type\cite{Bargmann49} that give  rise to the same scattering length and
 effective range. These potentials offer the unique advantage
 of having known analytic solutions so that the associated Jost functions
 can be constructed explicitly.

 A knowledge of  the Jost functions
 in turn determines  both the scattering and bound state properties of the
 $K\bar K$ system without futher approximation\cite{Newton56}.  In Section IV
 we carry out this program and compute both the mass and decay
 width of the kaonic molecule from the relevant Jost function that includes
 annihilation contributions. The resulting complex total energy for this system
 is $ (981-25i)$ MeV that is in reasonable agreement with the recent  experimental
 data from Fermilab\cite{fermilab2001}  that give
 $ [(975\pm 3)-(22\pm 2)i]$ MeV. 
 
 We also give computed elastic 
 and reaction cross sections  for $K\bar K$ scattering, the $\pi \pi $ inelasticities, and
 the cross section for the inverse process, $\pi^+\pi^-\to
 K\bar K$, for which data  exists, using detailed balance. 
Those  calculations disagree with the measured cross section, particularly near
threshold\cite{SDP73,ADM77,BHyams1973}.

 On the other hand the similarity in the pole position in both the recent Fermilab measurements as well as the calculated
position of this pole is striking. 
We also show that the molecular picture is totally inadequate for describing the 2 photon decay width of $f_0(980)$, which comes out to be an
order of magnitude larger than experiment. Taken together with the overall underestimate of the $\pi^+\pi^-\to K\bar K$ data, this suggests
that the $f_0(980)$ ground state cannot be a pure molecular state (see
 also\cite{Deser54}). This point is taken further at the end of Section \ref{secIV:numerical_results}.

 We take up the discussion of kaonium in Section V. Since kaonium is a
 mixture of isoscalar and isovector states due to isospin--breaking introduced by the
 Coulomb potential, we appeal to  the no--internal--mixing approximation
 that was introduced in connection with
  isobaric analog states in nuclei.  In this approximation
  one joins linear combinations of the $I=0$ and $I=1$ isospin amplitudes in the
  external region where only the Coulomb field is important, onto
  (in this case) known solutions of good isospin of the wave equation
  in the internal region where the strong field is dominant.  This allows
  one to
   construct the Jost function for kaonium without further approximation.
  Only momenta of
  ${\cal O}(M_K\alpha)$, where $\alpha\approx 1/137$,
   are  of interest for kaonium. This fact allows one to recast
   the condition that
  its bound states are given by the  zeros of the Jost function
    in the lower half of the complex  momentum plane as
     an eigenvalue equation of the Kudryavtsev--Popov type\cite{KP79},
 in which the strong field effects only enter through the $K^+K^-$
 scattering length.
  The solutions of this equation
  show that the kaonium levels are shifted upwards from their pure
  Coulomb values and acquire lifetimes between $10^{-18}$--$10^{-16}$ sec
  in the presence of the $K\bar K$ molecular ground state.

  We also show explicitly that the
  strong field introduces an additional node into  the  eigenfunctions
  of kaonium that lies at the $K^+K^- $
  scattering length. The kaonium levels are therefore to be viewed as excited states of the
  kaonic
  molecule in the combined Coulomb and  strong fields of the system. This in turn has the
  effect of enhancing the $K^+K^- \to 2\pi$  annihilation widths by 
  two orders 
   of magnitude over what they would have been under Coulomb binding of
  the charged kaon pair alone.

\section{formalism}
In order to obtain a non--relativistic equation for describing possible bound states
of  a $K\bar K$ meson pair resulting from  vector meson exchange  between them
we  briefly recall the derivation given by Landau and Lifshitz\cite{LL82}
that starts out with  the four--point vertex $i\Gamma[p_3,p_4;p_1,p_2]$
that enters the Bethe--Salpeter (BS) equation\footnote{The standard
graphical representation of the BS equation is given in Ref.~\cite{LL82}
to which we refer for further details.}.
The $p_1,p_2$  and $p_3,p_4$ are incoming and outgoing meson four momenta
 respectively.
 Then the homogeneous equation
for this vertex that determines the bound state poles of the BS equation
is
\begin{eqnarray}
i\Gamma[p_3,p_4;p_1,p_2]&=&
\int \frac{d^4 q}{(2\pi)^4}
i\hat\Gamma[p_3,p_4;q,p_3+p_4-q](-i)D(q)(-i)D(p_3+p_4-q)
i\Gamma[q,p_3+p_4-q;p_1p_2]
\label{e:BS}
\end{eqnarray}
All momentum labels formally flow from right to left, and the sum of each
pair on either side of the semi--colon equals
the total four momentum $P$ of the incoming pair which is conserved throughout
the
diagrammatic equation.  The hatted vertex is the irreducible piece that
generates
$\Gamma$ by iteration. The $D's$ are meson propagators.
For a free meson of four momentum $q$ and mass $M$ one has
\begin{eqnarray}
-iD(q) =\frac{i}{q^2-M^2+i0} \approx (\frac{1}{2M})\frac{i}{q_0-M-{\bf
q}^2/2M+i0}
\label{e:NRP}
\end{eqnarray}
if in addition we move to the non--relativistic limit by formally suppressing propagation
``backwards'' in time,
i.e. by omitting the anti--particle pole in the upper half of the complex $q_0$ plane.
We note\cite{LL82}  that the momenta $p_1,p_2$ are simply
labels in (\ref{e:BS}) that are not determined by the equation at all. So
one can  simply drop them. The other simplification is to observe that the
combination
\begin{eqnarray}
\chi(p_3,p_4)=D(p_3)\Gamma[p_3,p_4] D(p_4)
\end{eqnarray}
appears under the integral sign. Thus  one can equally well recast
 Eq.~(\ref{e:BS}) as an integral equation for $\chi(p_3,p_4)$. This is most usefully
 written down in the center of mass (CM) system for which $[p_3,p_4]=\pm p+P/2$ with
 $p=[p_0,\bf p]$ and $P=[P_0,0]$, where $P_0$ is the total energy in the CM frame.
Then the equation determining the bound states reads\cite{LL82}
\begin{eqnarray}
&&i\chi(p,P)=D(p+\frac{1}{2}P)D(-p+\frac{1}{2}P)
\int
\frac{d^4q}{(2\pi)^4}\hat\Gamma[p+\frac{1}{2}P,-p+\frac{1}{2}P;q+\frac{1}{2}P,-q+\frac{1}{2}P]
 \chi(q,P)
\nonumber
\\
\label{e:waveq}
\end{eqnarray}
where $\chi(p_3,p_4)=\chi(p+P/2,-p+P/2)\equiv\chi(p,P)$.

To make further progress towards a non--relativistic equation, the
vertex $\hat\Gamma$ should not depend on the time components of the relative
outgoing and incoming four momenta $p=[p_0,\bf p]$ and $q=[q_0,\bf q]$, i.e.
\begin{eqnarray}
 \hat\Gamma[p+\frac{1}{2}P,-p+\frac{1}{2}P;q+\frac{1}{2}P,-q+\frac{1}{2}P]
 \approx \hat\Gamma[{\bf p},{\bf q},P_0]
 \end{eqnarray}
Should this be the case  one can then integrate out these time components
from $\chi$ to obtain what is effectively the $3D$ fourier
transform of the  wave function for relative motion,
\begin{eqnarray}
 &&\psi({\bf p})=\int_{-\infty}^{\infty}\frac{dp_0}{2\pi}\chi(p,P)
 \end{eqnarray}
Carrying out the integrals over the time components of the relative four momenta
one finally arrives at the desired equation,
\begin{eqnarray}
 (\frac{{\bf p^2}}{M}+2M-P_0)\psi({\bf p})-\int \frac{d^3{\bf q}}{(2\pi)^3}
 \frac{\hat\Gamma[{\bf p},{\bf q},P_0]}{4M^2}\psi({\bf q})=0
 \label{e:waveeqn}
 \end{eqnarray}
that makes use of the non--relativistic approximation
\begin{eqnarray}
\int_{-\infty}^{\infty}\frac{dp_0}{2\pi i}D(p+\frac{1}{2}P_0)
(D(-p+\frac{1}{2}P_0)
\approx\frac{1}{4M^2}\Big(\frac{1}{{\bf p^2}/M+2M-P_0}\Big)
\label{e:doubleint}
\end{eqnarray}
as given by Eq.~(\ref{e:NRP}).  One recognises Eq.~(\ref{e:waveeqn}) as the
Schroedinger wave equation in momentum space for relative motion in
a potential
\begin{eqnarray}
V({\bf p},{\bf q})=-\frac{\hat\Gamma[{\bf p},{\bf q},P_0]}{4M^2}
\label{e:potential}
\end{eqnarray}
and binding energy $\varepsilon=2M-P_0\ge 0$. Should
$\hat\Gamma[{\bf p},{\bf q},P_0]$ only depend on the difference $\bf{p-q}$, the corresponding
potential will be local in coordinate space.
Since $\hat\Gamma$ is a $relativistic$ proper vertex that
in lowest order
gives the $T$ matrix, the mass factor $1/4M^2$  that converts $\hat\Gamma$
into a non--relativistic potential is the same factor\cite{GLRG01} that relates the
relativistic $T$ matrix to its non--relativistic counterpart.

\section{One boson exchange potentials}
In the following we use
Eq.~(\ref{e:potential}) to investigate the possible binding of the
 $K\bar K$ system via one-boson exchange potentials by constructing $\hat\Gamma$
 from
 the relevant pieces of the  standard  $SU(3)_V\times SU(3)_A$ 
 invariant Lagrangian, the derivation and properties of which are  described
 in detail in Ref.~\cite{deSwart63}. For $K\bar K$ scattering the relevant
 interaction Lagrangians for our purposes are
 \begin{eqnarray}
 &&{\cal L}_{KK\rho}=g_{KK\rho}[K^\dagger \vec \tau \partial_\mu K]\vec\rho^\mu+cc
 \nonumber
 \\
 &&{\cal L}_{KK\omega}=g_{KK\omega}[K^\dagger \partial_\mu K]\omega^\mu+cc
 \nonumber
 \\
 &&{\cal L}_{KK\phi}=g_{KK\phi}[K^\dagger \partial_\mu K]\phi^\mu+cc
 \label{e:exchange}
 \end{eqnarray}
which, together with
\begin{eqnarray}
&&{\cal L}_{\pi KK^*}=g_{\pi KK^*}[\partial^\mu\vec\pi\cdot K^\dagger\vec \tau
K^*_\mu-\vec\pi\cdot\partial^\mu K^\dagger \vec\tau K^*_\mu]+cc
\label{e:annihil}
\end{eqnarray}
generate  interaction potentials for
$K\bar K\to K\bar K$ scattering via  $\rho,\omega,\phi$
vector meson exchange,
as well as for $K\bar K\to \pi\pi$ annihilation via
 $K^*(892)$ strange meson exchange.
 The $K's$
 are all isospin doublets and $cc$ stands for the additional charge conjugation
 contribution with $K\to\bar K$, $K^*\to\bar K^*$.

The  coupling constants in these expressions are all fixed in terms of
the $\rho\pi\pi$ coupling constant $g_{\rho\pi\pi}$ by $SU(3)$ 
symmetry relations\cite{deSwart63},
\begin{eqnarray}
&&g_{KK\rho}=g_{\bar K\bar K\rho}= \frac{1}{2}g_{\rho\pi\pi}
\nonumber
\\
&&g_{KK\omega}=g_{\bar K\bar K\omega}= -\frac{1}{2}g_{\rho\pi\pi}
\nonumber
\\
&&g_{KK\phi}=g_{\bar K\bar K\phi} =\frac{1}{\sqrt{2}}g_{\rho\pi\pi}
\nonumber
\\
&&g_{\pi K K^*}=g_{\pi \bar K \bar K^*}= -\frac{1}{2}g_{\rho\pi\pi}
\label{e:coupling}
\end{eqnarray}
On the other hand the $\rho\pi\pi$ coupling is determined by the KSRF
relation\cite{KSRF66} as
$g_{\rho\pi\pi}\approx  M_\rho/(\sqrt{2}f_\pi)\approx 6$  in terms of the $\rho$ meson mass
$M_\rho$ and the pion weak decay constant $f_\pi\approx 93$MeV.  In this sense, then,
there are no free parameters in the calculation of the exchange potentials.
They only contain physical meson masses and known coupling constants.

In the non--relativistic limit, $p_i+p_j\approx [M_i+M_j,0]$,
$p_i-p_j\approx [0,{\bf p}_i-{\bf p}_j]$
only the $t$--channel scattering diagrams are relevant
for determining $\hat\Gamma$. In this limit these amplitudes all have a
common Yukawa--like form
\begin{eqnarray}
\hat\Gamma[p+\frac{1}{2}P,-p+\frac{1}{2}P;q+\frac{1}{2}P,-q+\frac{1}{2}P]
\approx  g^2C_In_s\frac{(M_1+M_3)(M_2+M_4)}{M^2+{\bf k}^2}
\label{e:gammahat}
\end{eqnarray}
in momentum space, where ${\bf k}={\bf p}-{\bf q}$ is the $t$ channel $3$--momentum transfer,
and $M_i$ are the masses in the indicated entrance and exit channels; M is the mass of the
exchanged boson and $g$ the coupling constant.
The isospin and  boson identity
 factors  $C_I$ and $n_s$  in Eq.~(\ref{e:gammahat})
are given in\cite{Lohse90}. They are
$C_0=[3,1,1]$ and $C_1=[-1,1,1]$ and $n_s=1$ for $\rho,\omega,\phi$
 exchange in the $K\bar K\to K\bar K$ isoscalar and isovector
 channels. The corresponding values for $K^*$ exchange in the
  $K\bar K\to\pi\pi$  isoscalar and  isovector channels
  are $C_0=-\sqrt{6}$, $C_1=-2$, and $n_s=1/\sqrt 2$.

\subsection{$K\bar K$ exchange potentials}
Since $\hat\Gamma$ is only a function of the $3$--momentum transfer, the resulting
potentials in Eq.~(\ref{e:potential}) are attractive local Yukawa potentials
in coordinate space, and the determination of their
possible bound states is numerically straightforward. While not necessary for
convergence in the non--relativistic case,
 we, however, also include at each vertex in Eq.~(\ref{e:gammahat}) the
 form factor containing an arbitrary cutoff $\Lambda$
\begin{eqnarray}
F^t({\bf k})=\Big(\frac{2\Lambda^2-M^2}{2\Lambda^2+{\bf k}^2}\Big)^2
\end{eqnarray}
that was used in  \cite{Lohse90,Sassen01} to obtain convergence
  in scattering calculations based on the  relativistic $\hat\Gamma$ vertices.
>From Eq.~(\ref{e:potential}) the coordinate space non--relativistic
$K\bar K\to K\bar K$
potential associated with the exchange of boson $M$ is then
\begin{eqnarray}
V(r)= -g^2C_IU(M,r)
\end{eqnarray}
after division by $4M_K^2$ where $M_K$ is the kaon mass, and $U(M,r)$
is the fourier transform
\begin{eqnarray}
&&U(M,r)=\int \frac{d^3{\bf k}}{(2\pi)^3} \frac{[F^t({\bf k})]^2}{M^2+{\bf k}^2}e^{i\vec{k}\cdot\vec{r}}=
\nonumber
\\
&&\frac{1}{4\pi } \Big\{\frac{e^{-Mr}}{r} -\frac{e^{-\sqrt{2}\Lambda r}}{r}
\Big[1+\frac{1}{16}(11-\frac{M^2}{2\Lambda^2}+\frac{M^4}{4\Lambda^4})
(1-\frac{M^2}{2\Lambda^2})(\sqrt{2}\Lambda r)+
\nonumber
\\
&&\frac{1}{16}(3-\frac{M^2}{2\Lambda^2})(1-\frac{M^2}{2\Lambda^2})^2(\sqrt{2}\Lambda r)^2
+\frac{1}{48}(1-\frac{M^2}{2\Lambda^2})^3(\sqrt{2}\Lambda r)^3\Big]\Big\}
\end{eqnarray}

The resulting  $K\bar K$
potential due to $\rho,\omega,\phi$ exchange are attractive in both
isospin channels $I=0$ and $1$,
\begin{eqnarray}
V_{I=0}(r)=-g^2_{\rho\pi\pi}\Big\{\frac{3}{4}U(M_\rho,r)+\frac{1}{4}U(M_\omega,r)
+\frac{1}{2}U(M_\phi,r)\Big\}
\label{e:OBE}
\end{eqnarray}
and
\begin{eqnarray}
V_{I=1}(r)=-g^2_{\rho\pi\pi}\Big\{-\frac{1}{4}U(M_\rho,r)+\frac{1}{4}U(M_\omega,r)
+\frac{1}{2}U(M_\phi,r)\Big\}
\label{e:OBET}
\end{eqnarray}
but too weakly so in the latter channel due to an
almost complete cancellation between  $\rho$ and $\omega$ exchange
to support a bound state.
In deriving these expressions we have exploited the $SU(3)$ relations (\ref{e:coupling}) to express all
coupling constants in terms of $g_{\rho\pi\pi}$.

These potentials are plotted in
 Fig.~\ref{f:fig1} for a strong coupling constant
 $\alpha_s=g^2_{\rho\pi\pi}/4\pi=2.9$ and cutoff  $\Lambda=4$ GeV.
In principle $\Lambda$ can be different for each exchanged boson\cite{Lohse90}.
One  notes that the  $V_{I}$ are now finite at the origin, in
contrast to the sum of pure  Yukawa
potentials to which they  reduce as $\Lambda\to\infty$.

The
coordinate space version of the wave equation  Eq.~(\ref{e:waveeqn})
that describes the relative $s$--state motion  in these potentials
becomes the usual Schroedinger equation
\begin{eqnarray}
  [\bigtriangledown^2+k^2-M_KV_{I}(r)]\psi(r)=0
\label{e:schr}
\end{eqnarray}
with total CM energy $P_0=2M_K+k^2/M_K$.
We ignore the charged to neutral kaon
mass difference and work with an average kaon mass $M_K$.
By numerically constructing
the $k\to 0$ scattering solutions one easily obtains the scattering lengths
and effective ranges that enter into the   effective range expansion
of the $s$--wave phase shifts $\delta(k)$
\begin{eqnarray}
k\cot\delta(k)=-\frac{1}{a_I}+\frac{1}{2}r_I k^2 +\cdots
\label{e:effrange}
\end{eqnarray}
in the two channels as
\begin{eqnarray}
a_0=5.835M^{-1}_K ,\quad &&r_0=1.187M^{-1}_K \quad ({\rm isoscalar})
\nonumber
\\
a_1=-0.401M^{-1}_K ,\quad &&r_1=3.702M^{-1}_K \quad ({\rm isovector})
\label{e:numerical}
\end{eqnarray}
Thus only the isoscalar potential supports a bound state. Its binding energy
is calculated to be $\varepsilon=18.63$ MeV, placing the  total mass
  for the bound $K\bar K$ system at $P_0=2M_K-\varepsilon=973.4$ MeV.
 Fig.~\ref{f:fig2} illustrates the  sensitivity of $\varepsilon$, $a_0$ and $r_0$
 on the choice of cutoff for this channel. For $\Lambda\gtrsim 4$ GeV
 the values of $a_0$ and $r_0$  are only weakly dependent on
 $\Lambda$.

Even though the isovector potential does
not support a bound state the $a_0(980)$ can be fully explained by
this model as a threshold effect \cite{Jansen95}. Effects of the $a_0(980)$ thus are
included in the calculation though it should be noticed  that decays to
the $\pi\eta$-channel are not considered in the analytical model.

 \subsection{Shape--independent approach}
 The relatively small (on a hadronic scale) binding energy $\varepsilon/ M_K\sim 0.04$ in the isoscalar channel
 (and no binding at all in the isovector channel) indicates that
 an effective range approach should be applicable. This is confirmed by
 recalculating the binding energy from the
 effective range formula\cite{Newton56}
 $\varepsilon\approx\kappa^2/M_K=18.60$ MeV where
 $\kappa=-a=0.194 M_K$, see Eq.~(\ref{e:absv}) below.
 Consequently it is entirely sufficient to
 characterize the $K\bar K$ interaction at low energies in terms of a scattering length and an
 effective range: the actual shape of the potential is immaterial.

 We make use of this feature in the following to replace the strong potentials $V_{I}$
 in Eqns.(\ref{e:OBE}) and (\ref{e:OBET})
 by  analytic potentials  of the Bargmann type given below that are phase
 equivalent to them. Bargmann\cite{Bargmann49}
 (see also\cite{Newton60} for an in depth overview) has shown how to construct
  families of potentials  that give rise to a prescribed Jost
  function $f(k)$. If one chooses
  \begin{eqnarray}
  f(k)=\frac{k-ia}{k-ib}
 \label{e:jostfcn}
 \end{eqnarray}
  then the potential that leads to this Jost function can be
  determined. It is\footnote{This is perhaps the
  simplest choice out of an entire family of phase equivalent potentials
  (the case $c=2$ in the notation of\cite{Newton60}).}
  \cite{Bargmann49,Newton60},
   \begin{eqnarray}
 V_I (r)= -\frac{1}{M_K}\frac{8b^2}{b^2-a^2}\Big[\frac{e^{br}}{b-a}+
 \frac{e^{-br}}{b+a}\Big]^{-2}
 \label{e:bargpot}
 \end{eqnarray}
  For this potential the effective range
 expansion is exact:
\begin{eqnarray}
k\cot\delta(k)=\frac{ba}{b-a}+\frac{k^2}{b-a}
\label{e:bargeffrange}
\end{eqnarray}
This allows one to identify the scattering length and effective range as
\begin{eqnarray}
a_I=-\frac{b-a}{ba},\quad r_I=\frac{2}{b-a}
\label{e:bargar}
\end{eqnarray}
where the actual values of the parameters $a$ and $b$ will depend on the
isospin channel.

Fixing  $a$ and $b$ is obviously equivalent to prescribing the
scattering length and effective range.
If $a$ is negative, $a=-\kappa$, say, then $f(k)$ has a zero at $k=-i\kappa$
in the lower half of the
complex $k$ plane, and there is a bound state\cite{Newton60} of binding energy
 $\varepsilon_s=\kappa^2/M_K$. Both the scattering and bound state wave
 functions of the Bargmann potential in Eq.~(\ref{e:bargpot}) are
 known explicitly. Further details will be found in the Appendix.
 The availability of such analytic solutions
 will be important for including the effects of the
 $K\bar K\to\pi\pi$  decay channel  non--perturbatively as discussed below.

Now invert Eq.~(\ref{e:bargar}) to construct  phase equivalent
potentials
that have the same scattering length
and effective range as
the original  exchange potentials in Eqs.~(\ref{e:OBE}) and
(\ref{e:OBET}). The required values of $a$ and $ b$ are
\begin{eqnarray}
 a=-&&\frac{1}{r_I}\Big[1-\sqrt{1-2r_I/a_I}\Big]
 \nonumber
 \\
 b=&&\frac{1}{r_I}\Big[1+\sqrt{1-2r_I/a_I}\Big]
 \label{e:ab}
 \end{eqnarray}
  or
  \begin{eqnarray}
  a= -0.1936{\rm M_K}\quad &&b= 1.491 {\rm M_K}\quad ({\rm isoscalar})
  \nonumber\\
  a= +0.9219{\rm M_K}\quad &&b= 1.462 {\rm M_K}\quad ({\rm isovector })
 \label{e:absv}
 \end{eqnarray}
 for the two channels in question. These parameters are
 summarized in  Table~\ref{t:table1} for easy reference.
The resulting set of phase equivalent potentials are included
in Fig.~\ref{f:fig1}.

\subsection{ $K\bar K$ annihilation potential}
>From Eq.~(\ref{e:gammahat}) one  reads off the irreducible
vertex for $K\bar K\to\pi\pi$ via $K^*$ exchange, as
\begin{eqnarray}
\hat\Gamma(K\bar K\to \pi\pi)\approx -C_I n_s g^2_{\pi K K^*}
\frac{(M_\pi+M_K)^2}{M_{K^*}^2}
\label{e:gammaKKpipi}
\end{eqnarray}
where $n_sC_I=-\sqrt 3$ or $-\sqrt 2$ in the isoscalar or isovector
channel. We have neglected the $3$--momentum transfer relative to the
large $K^*(892)$ mass.  As shown in Fig.~\ref{f:fig3}, this amounts to
introducing point vertices that can contribute to $K\bar K$ scattering via an
intermediate pion loop. The contribution from this
interaction is complex due to the possibility of on--shell decays $K\bar K\to\pi\pi$
in the intermediate state, and the  resulting $K\bar K$ bound state acquires a width.
The details are as follows: calling the contribution from Fig.~\ref{f:fig3}
$i\hat\Gamma^{(1)}$ one finds
\begin{eqnarray}
 i\hat\Gamma^{(1)} &&=i\hat\Gamma^2(K \bar K\to \pi\pi)
 \int\frac{d^3{\bf l}}{(2\pi)^3}\int _{-\infty}^\infty\frac{dl_0}{2\pi i}
 D(l+\frac{1}{2}P_0)D(-l+\frac{1}{2}P_0)
\nonumber\\
&&=i\hat\Gamma^2(K \bar K\to \pi\pi) \frac{1}{4M^2_\pi}\int \frac{d^3{\bf l}}{(2\pi)^3}
 \frac{1}{{\bf l}^2/M_\pi+2M_\pi-P_0}
\nonumber
\\
&&=i\hat\Gamma^2(K \bar K\to \pi\pi) \frac{J(P_0)}{4M^2_\pi}
\label{e:gammahat1}
\end{eqnarray}
in view of Eq.~(\ref{e:doubleint}).   As summarized briefly in Eqs.~(\ref{e:d}) and
(\ref{e:JP}) of the Appendix the
three dimensional integral $J(P_0)$ can be dimensionally regulated
in $d$ dimensions without difficulty to obtain a finite result\cite{GLRG01}
for $d\to3$. The associated annihilation
 potential $V_{K\bar K}$ is then obtained from Eq.~(\ref{e:potential}) as an
attractive contact potential in coordinate space,
\begin{eqnarray}
V_{K\bar K}(r)=-\frac{\hat\Gamma^2(K \bar K\to \pi\pi)}{(4M^2_K)(4M^2_\pi)}J(P_0)
\delta^3({\bf r})=-c^2_0\delta^3({\bf r})
\label{e:vcontact}
\end{eqnarray}
with
\begin{eqnarray}
c^2_0=
\frac{3}{16}\frac{(g^2_{\pi K K^*})^2}{M_\pi^2M_K^2}
\Big(\frac{M_\pi+M_K}{M_{K^*}}\Big)^4J(P_0)\quad {(\rm isoscalar})
\label{e:c0}
\end{eqnarray}
The value of this factor in the isovector channel is $c_1^2=\frac{2}{3}c_0^2$.
The function
\begin{eqnarray}
J(P_0)=i\frac{M_\pi}{4\pi}\Big[M_\pi(P_0-2M_\pi)\Big]^{1/2}
\label{e:JP0}
\end{eqnarray}
 is a complex function of $P_0$ that is positive and pure imaginary along
the upper lip of the branch cut
 extending  from $2M_\pi$ to $\infty$
along the real $P_0$ axis, see Eq.~(\ref{e:JP}).
\subsection{Jost function for a  finite  plus delta function
potential at the origin}
We investigate the influence of the  annihilation potential
on the properties of the $K\bar K$ system
 by constructing the revised Jost function for the sum
 $V_I+V_{K\bar K}$. Since  $V_{K\bar K}$ is a
contact interaction, and the solutions in the  potential $V_I$ are known,
this can be done without approximation.
The new scattering problem to be solved reads
\begin{eqnarray}
  [\bigtriangledown^2+k^2-M_KV_{I}(r)+M_Kc^2_I\delta^3({\bf r})]\psi(r)=0
\label{e:newschr}
\end{eqnarray}
 Consider isospin $I=0$. The presence of the delta function potential obliges one to
give up the
boundary condition $\varphi(k,r)\sim r\to 0$ on the  scattering wave
function $\varphi(k,r)=r\psi(k,r)$ at the origin in favor of
 prescribing\footnote{
  This can also be interpreted as  $\varphi(k,r)$  behaving
  discontinuously at the origin,
   $\varphi(k,0)=0$, but
  lim $\varphi(k,r)\neq 0$ as $r\to 0$ under the influence of the delta function
  potential.}
its logarithmic derivative  $\varphi^\prime/\varphi$  there\cite{Newton60}.
By integrating Eq.~(\ref{e:newschr}) over the volume of an infinitesimal sphere
centered at the origin, one finds that the logarithmic derivative at the
origin
is fixed by the strength of the delta potential according to
\begin{eqnarray}
\frac{\varphi^\prime}{\varphi}=\frac{4\pi}{M_Kc^2_0}=\zeta
\label{e:zeta}
\end{eqnarray}
  The point is now that for $r>0$ both the boundary condition
  on $\varphi(k,r)$ and  the closed form of irregular solution of
  Eq.~(\ref{e:newschr}) are known explicitly: they are given by Eqs.~(\ref{e:zeta})
  and (\ref{e:irreg}) respectively.  By constructing the Wronskian of
  this pair
  in the limit $r\to 0$ one obtains
  the revised Jost function as
  \begin{eqnarray}
   f(k)=W[f,\varphi]=
    \Big[(\zeta+ik)\frac{k-ia}{k-ib}+i\frac{b^2-a^2}{k-ib}\Big]
   \label{e:revjost}
   \end{eqnarray}
 after setting lim $\varphi(k,r)\to 1$  for convenience.
This choice is unimportant since neither the root of $f(k)$ nor the scattering
phase shift depends on it. The value of the logarithmic derivative $\zeta$
is $k$--dependent through $P_0$: from Eqs.~(\ref{e:c0}) and  (\ref{e:zeta})  one finds
\begin{eqnarray}
  &&\zeta(k)=-\frac{256i}{3\alpha^2_s}M_\pi M_K
  \Big(\frac{M_{K^*}}{M_K+M_\pi}\Big)^4
  [M_\pi(P_0-2M_\pi)]^{-1/2},
  \nonumber
  \\
  &&P_0=2M_K+\frac{k^2}{M_K}
  \label{e:zetak}
  \end{eqnarray}
  after taking the $SU(3)$ value $g_{\pi K K^*}=-g_{\rho\pi\pi}/2$ for the coupling
  constant.
 \section{Numerical results}
\label{secIV:numerical_results}
 The single equation (\ref{e:revjost})  contains all the  necessary
  information on the scattering and bound states of the
  two--body $K\bar K$ system. Moreover, this
  information can be extracted without further approximation. We use this form
 of $f(k)$ to identify the revised scattering length and effective range as well as
 the bound state energy in the presence of $K\bar K\to 2\pi$ annihilation.

\subsubsection{Effective range expansion}
  The $K\bar K\to \pi\pi$ annihilation channel renders the scattering matrix
  $S(k)=f(k)/f(-k)$ non--unitary, since now $f^*(k)\neq f(-k)$.
  This can be  seen directly from Eq.~(\ref{e:revjost}) after noting that
  the logarithmic derivative $\zeta$ is even in $k$ and purely negative imaginary
   for real $k$.  The  scattering phase shift is now complex.
   In order to identify the scattering length and effective range
   in the presence of absorption, one can either maintain the
   expansion (\ref{e:effrange}), thereby introducing complex
   effective range parameters\cite{Deser54,KP79}, or define\cite{Petersen71}
   these in terms
   of the real part $\delta(k)$
    of the total phase shift (not the phase of $S(k)$),
   \begin{eqnarray}
    S(k)=f(k)/f(-k)=\eta e^{2i\delta(k)},\quad 0<\eta<1
    \label{e:smat}
    \end{eqnarray}
   In the latter case
 \begin{eqnarray}
  k\cot\delta(k)=ik\frac{S(k)+\eta}{S(k)-\eta}=ik\frac{f(k)|f(-k)|+f(-k)|f(k)|}
 {f(k)|f(-k)|-f(-k)|f(k)|}
 \end{eqnarray}
 that shows explicitly that $k\cot\delta(k)$ is an even function of $k$.
 Set $\zeta(k)=-i\xi(k)$ in Eq.~(\ref{e:revjost}) for the Jost function,
 and then expand the right hand
 side to ${\cal O}(k^2)$. After some calculation
 one establishes the revised  effective range expansion
 \begin{eqnarray}
 k\cot\delta(k)=-\frac{1}{a^\prime_0}+\frac{1}{2}r^\prime_0 k^2+\cdots
 \label{e:neweffrange}
 \end{eqnarray}
 with  the exact expressions
 \begin{eqnarray}
 &&-\frac{1}{a^\prime_0} = \frac{b}{b-a}\Big\{\frac{a^2\xi^2+(b^2-a^2)^2}{a\xi^2-2ab^2+a^3-b^3}\Big\}
 \label{e:a0prime}
 \end{eqnarray}
 and
 \begin{eqnarray}
 &&\frac{1}{2}r^\prime_0= \frac{a^2(b^2-a^2)^4+a(b+a)[a^3(b-a)(5b^2-3a^2)+
 b^4(a^2-2ab+2b^2)]\xi^2}
 {(b-a)[a^2\xi^2+(b^2-a^2)^2][a^3-b^3-2ab^2+a\xi^2]^2}+
 \nonumber\\
 &&\frac{a(b+a)[a(b-a)(b^2-3a^2)-b^4]\xi^4+a^4\xi^6}
 {(b-a)[a^2\xi^2+(b^2-a^2)^2][a^3-b^3-2ab^2+a\xi^2]^2}-
 \nonumber\\
 &&2\xi\xi^\prime ab^4\frac{[(a^2+ab-b^2)(a^3-ab^2-b^3)+a^2\xi^2(b+a)]}
 {(b-a)[a^2\xi^2+(b^2-a^2)^2][a^3-b^3-2ab^2+a\xi^2]^2}
 \label{e:r0prime}
 \end{eqnarray}
 for the new inverse scattering length and effective range.
 Here
 \begin{eqnarray}
&&\xi=i\zeta= \frac{256}{3\alpha^2_s}M_\pi M_K
  \Big(\frac{M_{K^*}}{M_K+M_\pi}\Big)^4
  [M_\pi(2M_K-2M_\pi)]^{-1/2} =17.409M_K
\nonumber\\
&&\xi^\prime=
  i\frac{\partial\zeta}{\partial k^2}=-\frac{128}{3\alpha^2_s}\sqrt{M_\pi}
  \Big(\frac{M_{K^*}}{M_K+M_\pi}\Big)^4
  [(2M_K-2M_\pi)]^{-3/2} =-6.064M^{-1}_K
 \label{e:xixipr}
 \end{eqnarray}
 at $k=0$,
 and the primes serve to
 distinguish  these scattering lengths and ranges from those appearing
 in Eq.~(\ref{e:bargeffrange}), to which they reduce in the
 no annihilation limit $\xi\to \infty$.
 The numerical values of $\xi$
 and
 $\xi^\prime$  have been calculated for
 meson masses $M_\pi=140$ MeV, $M_K=496$ MeV, $M_{K^*}=892$ MeV
 and  coupling constant $\alpha_s=2.9$. Using these
  together with the values for $a$ and $b$ given in Table~\ref{t:table1},
  one obtains
  \begin{eqnarray}
  a^\prime_0=4.281M^{-1}_K;\quad r^\prime_0=3.203M^{-1}_K \quad ({\rm isoscalar})
  \label{e:revnumerical}
  \end{eqnarray}
 as predictions for the $K\bar K$ isoscalar scattering  length and effective range
 in the presence of absorption. The corresponding values for the isovector
 channel are still
  \begin{eqnarray}
  a^\prime_1=-0.401M^{-1}_K;\quad r^\prime_1=3.702M^{-1}_K
  \quad ({\rm isovector})
  \label{e:revnumericaltr}
  \end{eqnarray}
since there is no annihilation contribution from the $l=0,\;I=1$  partial
wave  to the $K\bar K$ scattering amplitude due to the identity of the
outgoing pions in the isospin representation.  Higher partial waves cannot contribute in
any event due to the contact nature of the annihilation interaction. 
In fact the $a_0(980)\to 2\pi$ decay channel is forbidden by $G$--parity.
Thus 
the  isovector
scattering length remains real if, as here, the $\pi\eta$ decays are ignored.

 \subsubsection{Bound states for $I=0$}
 The bound states of Eq.~(\ref{e:newschr}) are determined\cite{Newton60}
 by the
 complex root(s) of $f(k)$ in Eq.~(\ref{e:revjost}), with Im$k<0$, i.e.
\begin {eqnarray}
(\zeta(k)+ik)(k-ia)+i(b^2-a^2)=0
\label{e:rootjost}
\end{eqnarray}
Let $k=p-i\kappa$, $\kappa>0$ be such a root. Then the total mass $M$ and decay width
$\Gamma$ of a bound $K\bar K$ pair at rest reads
    \begin{eqnarray}
    P_0=M-\frac{i}{2}\Gamma=2M_k+\frac{(p-i\kappa)^2}{M_K}
    \label{e:P0complex}
    \end{eqnarray}

   This root can only be found numerically as will be done
   shortly\footnote{For the values of $a$ and $b$  and the logarithmic
   derivatives given in Table~\ref{t:table1}
   it turns out that
   there is only one such root
   in the isoscalar channel,
   and none in the isovector channel.}.
   However, since the magnitude of the  absorptive coupling
   $M^2_K|c^2_0|\sim 4\pi M_K/\xi\sim 0.7$ is quite small relative to
    the  isoscalar potential strength, a
   perturbative solution in inverse powers of $\zeta$ is useful.
   Replace $\zeta$ by its value $-i\xi$ at $k=0$ (more correctly
   $k\approx -i\kappa$
   but the difference is quantitatively insignificant).
    Then Eq.~(\ref{e:rootjost}) reduces to  a quadratic equation for
   $k$. The relevant root is
   \begin{eqnarray}
   k=  \frac{b^2-\kappa^2}{\xi}-i\kappa+O(\xi^{-2}) \approx (0.1255-0.1936i)M_K
   \label{e:approxk}
   \end{eqnarray}
   where $k=ia=-i\kappa=-0.1936M_Ki$ is the root leading to the bound state
    in the absence of absorption, and $b=1.491M_K$.  Working to the
    same order in $\xi^{-1}$ (i.e. neglecting the real part of $k$),
    one finds  that
     \begin{eqnarray}
    P_0\approx 2M_K-\frac{\kappa^2 }{M_K}-2i\kappa\frac{(b^2-\kappa^2)}{M_K\xi}
    \label{e:P0perturb}
    \end{eqnarray}
    Since $4\pi\psi^2(0)=2\kappa(b^2-\kappa^2)$  from Eq.~(\ref{e:bargu})
     the result in Eq.~(\ref{e:P0perturb}) is equivalent to treating
      the annihilation potential  $-c^2_0\delta^3({\bf r})$
     as a first order perturbation of $P_0$,
     \begin{eqnarray}
    P_0\approx 2M_K-\frac{\kappa^2 }{M_K}-\frac{i}{2}\Gamma;
     \quad \Gamma=-2ic^2_0\psi^2(0)=\frac{8\pi}{M_K\xi}\psi^2(0)
     \label{e:P0perturb1}
    \end{eqnarray}
     Upon recalling that $-i\xi^{-1}\sim -J(P_0)$, one sees from this form
     that the complex solution for $P_0$ always lies
    on the $second$ sheet $-2\pi\leq \phi\leq 0$ of the cut complex $P_0$ plane
    for $J(P_0)$. Otherwise the signs of the imaginary parts of $P_0$ and
    $-J(P_0)$ will differ, and no solution is possible.
    Taking cognizance of this fact
    the exact root of Eq.~(\ref{e:rootjost}) and resulting
    value for $P_0$ are found numerically to be
    \begin{eqnarray}
    k=(0.1254-0.1972i)M_K;\quad P_0=(981-25i){\rm MeV}
    \label{e:exactroot}
    \end{eqnarray}
     using the  the previously determined values of $a$ and
      $b$ shown in Table~\ref{t:table1}.  Notice that the exact value of $k$ lies
      rather close to the perturbation estimate given  by Eq.~(\ref{e:approxk})

 The exact numerical results for the Bargmann potential that is
 phase--equivalent to
 Eq.~(\ref{e:OBE})  are given in the
 first row of Table~\ref{t:table2}.  We contrast these results with the mass and
 width  calculations based on
  the Weinstein--Isgur non--relativistic potential model\cite{WS90}, as well
 as the predictions of Morgan and Pennington\cite{MP93}, based on a parametrization
 of the Jost function{\footnote{Note: these authors use the opposite convention
 for the definition of the Jost function to that employed in\cite{Newton60}.
 Thus their sheet II for complex $k$ corresponds
 to Im $k\leq 0$ in our notation.} with parameters determined from experimental
 phase shifts.  These calculations are also compared with the earlier experimental
 results summarized by
the Particle Data Group (PDG)\cite{PDG92}, as well as the recent
data from the Fermilab E791 Collaboration\cite{fermilab2001}.

In order to extract
Bargmann potential parameters that would
 lead to the  mass and width values of either Weinstein--Isgur,
 or Morgan and Pennington, we have turned
  Eq.~(\ref{e:rootjost}) around and asked what values of
     $a$ and $b$  produce a root  $k=p-i\kappa$ that gives a prescribed
     value  of the
     complex total energy in Eq.~(\ref{e:P0complex}). The answer is
 \begin{eqnarray}
 &&a=-\kappa-p\frac{\nu+\kappa}{p-\xi}
 \\
 &&b=\sqrt{a^2-p(p-\xi)+(\nu+\kappa)(a+\kappa)}
 \label{e:findab}
 \end{eqnarray}
after setting $\zeta(p-i\kappa)=\nu-i\xi$.  We have treated the experimental data
in the same way.
These relations  allow one to  establish a direct link between the mass and
decay width
of the bound  $K\bar K$ state,  and thence extract the isoscalar  scattering length and effective
range  via Eqs.~(\ref{e:a0prime})
and (\ref{e:r0prime}), or Eqs.~(\ref{e:cmplxln}) and (\ref{e:cmplxr}).

The  effective range expansion parameters  obtained in this
way for the the Weinstein--Isgur calculations, the Morgan--Pennington  fits,
 the PDG data, and the Fermilab data, are also shown in Table~\ref{t:table2}.
Collectively these results are all internally consistent, and suggest that the 
non--relativistic  OBE potential
that includes absorption via $K^*$ exchange, provides a natural
description for the properties  of the $f_0(980)$ scalar meson
 as a bound $K\bar K$ pair. In particular, the prediction of the correct
  decay width from $K^*$ exchange, which is parameter--free
 and can be treated without approximation, adds weight to this
 interpretation. In this connection it is also interesting to compare with the
 current--algebra prediction\cite{RHLRT2002} of $\Gamma\approx 660$ MeV for
 the much larger
 decay width ($600-1000$ MeV, also dominantly into two pions) of the
 $f_0(400-1200)$ or $\sigma$ scalar meson, that is considered as
 a candidate for $q\bar q$ bound state. 
 
 \subsubsection{Two--photon decay of $f_0(980)$}
 The two--photon decay of the $f_0(980)$, considered as a bound kaon--antikaon 
 pair, can be treated by the same method. Working in a suitable gauge, 
 the two--photon annihilation potential $V_{\gamma\gamma}$ is given by the single diagram in 
 Fig.~\ref{f:fig3} with the 
  $K\bar K\to 2\pi$ vertices and pion propagators replaced by the analagous
electromagnetic  $K^+K^-\to 2\gamma$ (''seagull'') vertices and photon 
propagators, after division by two to
compensate for crossing. One obtains
\begin{eqnarray} 
V_{\gamma\gamma}=-c^2_{\gamma\gamma}\delta^3({\bf r})
\end{eqnarray}
Without counterterms,  $c^2_{\gamma\gamma}$ is formally divergent due to the
photon loop. However, the imaginary part is finite,
\begin{eqnarray}
{\rm Im}c^2_{\gamma\gamma} =\frac{1}{2}(\frac{2\pi\alpha^2}{M^2_K})
\end{eqnarray}
where  $\alpha= e^2/4\pi\approx 1/137$ is the fine structure constant. 
The two--photon decay
width then follows from the second expression  in 
Eq.~(\ref{e:P0perturb1}) as
\begin{eqnarray}
\Gamma_{\gamma\gamma}=2{\rm Im}c^2_{\gamma\gamma}\psi^2_{K^+K^-} =
\frac{2\pi\alpha^2}{M^2_K}\psi^2_{K^+K^-} 
=\frac{\alpha^2 \kappa(b^2-\kappa^2)}{2M^2_K} = 5.59 {\rm keV}
 \label{e:2gammawidth}
 \end{eqnarray}

where $\psi_{K^+K^-}=\psi<K^+K^-|K\bar K>$ is 
the probability amplitude for  
finding a $K^+K^-$ pair in the $f_0(980)$ ground state. 
For a state of 
good isospin, $<K^+K^-|K\bar K>\,\, =1/\sqrt{2}$.
One can verify this result independently  by noting that
the coefficient  $2\pi\alpha^2/M^2_K =v\sigma_{\gamma\gamma}$, where 
$\sigma_{\gamma\gamma}$ gives the low velocity cross section for the 
annihilation of free,
 charged boson--antiboson pairs of relative velocity $v$ into two photons\cite{IZ80}.
The decay width of the bound state is then calculated in the standard 
way\cite{IZ80}  as
$\sigma_{\gamma\gamma}$ times the flux density, or 
$\sigma_{\gamma\gamma}\times
v\psi^2_{K^+K^-}$, which just reproduces Eq.~(\ref{e:2gammawidth}).
 
The numerical value found above for $\Gamma_{\gamma\gamma}$
is an order of magnitude larger than both the estimate given in 
\cite{Barnes1985} and also  the
currently quoted 
value  
of $0.39^{+0.10}_{-0.13}$ keV  from experiment\cite{PRD96}. 
In the molecular picture this discrepancy comes about because of the extreme sensitivity of the spatial
wave function at the origin to the assumed binding energy of the $K\bar K$ pair. In 
\cite{Barnes1985} the author estimates $\psi(0)\approx \frac{1}{10} M^{3/2}_K$ using
a gaussian potential fitted to a binding energy of $10$ MeV. 
  This is a factor $\sim 3$ smaller than our estimate of $\psi(0)$
that corresponds to a binding energy of $18.6$ MeV, which in turn 
increases our estimated width by an order of magnitude over that given 
in\cite{Barnes1985} (see Fig.~\ref{f:fig8}).  

On the other hand  
 the $2\pi$ decay width, which  is also proportional to $\psi^2(0)$, 
 is correctly reproduced by our wave function. 
Thus it would not seem possible to reproduce both the measured pole position of the
$f_0(980)$ decaying state in the complex plane and at the same time the two photon decay width in the molecular 
picture without intoducing some additional mechanism that reduces the  
 fractional parentage $f_p=\,\, <K^+K^-|K\bar K>$ of $K^+K^-$ in the $f_0(980)$ ground state.
For, apart from this factor, 
 the branching ratio of electromagnetic to hadronic decays is  given by
 the expression
 \begin{eqnarray}
 \frac{\Gamma_{\gamma\gamma}}{\Gamma} 
 \approx\frac{\alpha^2\xi}{8M_K}|f_p|^2=\frac{2}{3\sqrt 2}\Big(\frac{e^2}{g^2_{\pi KK^*}}\Big)^2 
 \Big(\frac{M_{K^*}}{M_\pi+M_K}\Big)^4
 \Big(\frac{M_\pi}{M_K-M_\pi}\Big)^{1/2}|f_p|^2
\approx 1.2\times 10^{-4} |f_p|^2
\end{eqnarray}
that contains known physical constants if one adopts the $SU(3)$ value $g_{\pi KK^*}=-1/2g_{\rho\pi\pi}$ for 
the $K\to \pi K^*$ coupling constant again. The only less well--determined
constant in this expression is $g_{\rho\pi\pi}$. However, the spread between the
various estimates\cite{Lohse90,Jansen95} of this coupling is not nearly large enough to cause the branching ratio
to alter by an order of magnitude. This, then leaves $f_p$. Since $\Gamma_{\gamma\gamma}/\Gamma\sim 10^{-5}$ experimentally, 
one would require  $f_p\sim
1/\sqrt{10}$ down from $1/\sqrt{2}$, suggesting that the assumption of a 
pure 
$K\bar K$ molecular ground state of good isospin is oversimplified.

A competing picture for the structure of $f_0(980)$ as a four--quark state 
has been 
developed in\cite{achasov82}, which  leads to an  estimate 
$\Gamma(f_0(980)\to \gamma\gamma)\approx 0.27$ keV \cite{Achasov82} that is also 
(with a particular choice of parameters)
not in contradiction 
with experiment.  One has therefore to conclude that agreement or otherwise 
between the
calculated and experimental  $2\gamma$ decay width of $f_0(980)$  by itself
is too ambiguous for deciding between a molecular versus a 
 $q^2\bar q^2$ four--quark  structure of this meson, and that some mixture of these 
two extreme pictures is probably indicated.

 \subsubsection{$K\bar K$ cross sections and complex scattering lengths}
 
   The linear combination of scattering amplitudes of 
  isospin zero and one, $f_p=(f_0+f_1)/2=(S_p-1)/2ik$  
  determines the $s$--wave  $S$ matrix as $S_p=(S_0+S_1)/2$ for
  the scattering and 
 reaction cross sections for $K\bar K$  scattering  
 in the absence of Coulomb interactions. 
 One has  
 \begin{eqnarray}
 \sigma_{elas}=\frac{\pi}{k^2}|1-S_p|^2, \quad
 \sigma_{r}=\frac{\pi}{ k^2}(1-|S_p|^2)
 \label{e:physicalsig}
 \end{eqnarray}
 where $k$  is the relative momentum of the colliding pair. 
 The reaction cross section has in turn  contributions from the 
 (quasi--elastic) ``charge transfer'' channel,  $K^+K^-\leftrightarrow K^0\bar K^0$,
 in addition to the two--pion annihilation channel. One finds $\sigma_{r}= \sigma_{ex}+
 \sigma_{a}$, with
 \begin{eqnarray}
 \sigma_{ex}=\frac{\pi}{k^2}|(\frac{S_0-S_1}{2})|^2,\quad 
 \sigma_a=\frac{1}{2}(\sigma^{(0)}_a+\sigma^{(1)}_a)
\label{e:sigex}
\end{eqnarray} 
since the annihilation cross section $\sigma_a$ is  
simply  the average 
of the corresponding annihilation cross sections 
\begin{eqnarray}
\sigma^{(I)}_a=\frac{\pi}{k^2}(1-|S_I|^2) 
\label{e:sigabsi}
\end{eqnarray}
in the separate isospin 
channels that add incoherently. 

The S--matrix elements of good isospin, $S_I$,  appearing in these expressions are
 given directly by Eq.~(\ref{e:smat})  in terms
 of the Jost functions for each isospin channel as 
 \begin{eqnarray}
 S_I(k)= \Big[\Big(\frac{(\xi-k)(k-ia)-(b^2-a^2)}{(\xi+k)(k+ia)+(b^2-a^2)}\Big)
 \Big(\frac{k+ib}{k-ib}\Big)\Big]_I
 \label{e:SI}
 \end{eqnarray}
 where the parameters in this expression refer to a specific isospin as indicated.
Then
\begin{eqnarray}
\sigma^{(I)}_{elas}=\frac{4\pi}{b^2+k^2}\Big[\frac{(b^2-a^2+ba+k^2)^2+(b-a)^2\xi^2}
{a^2(\xi+k)^2+(b^2-a^2+k(\xi+k))^2}\Big]
\nonumber\\
\sigma^{(I)}_a=\frac{4\pi\xi}{k}\Big[\frac{b^2+k^2}
{a^2(\xi+k)^2+(b^2-a^2+k(\xi+k))^2}\Big]
\label{e:sigI}
\end{eqnarray}
Note that $\sigma^{(I)}_a\sim 1/k$ at small $k$ as befits a true absorption 
cross section.
One can confirm the  expression for the annihilation cross
section $\sigma^{(I)}_a$ by considering the flux loss in  each 
isospin channel out of the 
  radial wave function 
$\varphi(k,r)/f(-k)$ for scattering, into an infinitesimal sphere located at the
origin. Employing the  boundary conditions $\varphi(k,r)\to 1$,
$\varphi^\prime(k,r)\to\zeta$ as $r\to 0$ used to obtain the 
Jost function $f(k)$ of Eq.~({\ref{e:revjost}), this flux loss is given without approximation 
by $8\pi/(M_K|f(-k)|^2)$.
 This is identical with the right hand 
side of the second of 
Eqs.~(\ref{e:sigI}) after division by the incident flux $v=2k/M_K$. 

 Finally, taking the first option 
 mentioned above Eq.~(\ref{e:smat}) and expanding 
 expanding $k\cot \delta_I(k)$, where $\delta_I$ is the phase of 
 $S_I(k)=exp(2i\delta_I)$, one has 
 \begin{eqnarray}
 k\cot\delta_I(k)=-1/a^{eff}_I+\frac{1}{2}r^{eff}_Ik^2+\cdots
\end{eqnarray}
  where the complex scattering length and effective range are given by the 
  much simpler expressions
\begin{eqnarray}
 a^{eff}_I
 &=&\big[\frac{b^2-a^2+ba-i(b-a)\xi}{b(b^2-a^2+ia\xi)}\Big]
 \label{e:cmplxln}
 \end{eqnarray} 
  and
  \begin{eqnarray}
  r^{eff}_I&=&-\frac{2(b-a)(\xi+ia)^2 }{[b^2-a^2+ba-i(b-a)\xi)]^2}
\label{e:cmplxr}
  \end{eqnarray}
  Values of these quantities in the $I=0$ and $I=1$ channels are listed in 
  Table~\ref{t:table3}.  Note that for $I=1$, these expressions reduce to
those in Eq.~(\ref{e:bargar}). The Re$(a^{eff}_I)$  of course coincide with
  the previous results, Eq.~(\ref{e:a0prime}), for the scattering lengths 
  when annihilation is present.
   From Eqs.~(\ref{e:physicalsig}) to (\ref{e:sigabsi}) the zero momentum
   elastic, charge transfer, and annihilation cross sections are given by
 \begin{eqnarray}
 \sigma_{elas}=4\pi|a^{eff}_p|^2,\quad 
 \sigma_{ex}=4\pi |a^{eff}_n|^2,\quad 
 \sigma_{a}=-\frac{4\pi}{k} Im(a^{eff}_p)
 \label{e:sigmas}
 \end{eqnarray}
 in the effective range approximation in terms of 
 $a^{eff}_p=(a^{eff}_0+a^{eff}_1)/2$,  
  the  common $K^+K^-$ or $K^0\bar K^0$ 
  scattering length in the absence of Coulomb interactions, 
  and $a^{eff}_n=(a^{eff}_0-a^{eff}_1)/2$.  
   
  One can show from Eq.~(\ref{e:cmplxln}) that the sign of Re$(a^{eff}_I)$ 
  is determined by the sign of the ``bare'' scattering length $a_I$ given by 
  Eq.~(\ref{e:bargar}) in the absence of absorption,
   and hence the sign of the parameter $a$ of the Bargmann potential
  that controls the existence $(a < 0)$ or not $(a > 0)$ of a bound state 
  in channel $I$; 
   Im$(a^{eff}_I)$ is independent of this sign.  
   Fig.~\ref{f:fig4} illustrates 
   how  the real and imaginary parts
   of $a^{eff}_p$ would vary  
   should this parameter vary (i.e. the binding energy vary)
   in the isoscalar channel. The remaining 
   parameters for this illustration have been kept the same as in 
   Table~\ref{t:table1}. 
   However, the effects for small variations in the binding energy from real  
   to virtual around zero are not
   large: for example by simply switching the sign of $a$ to $+0.9136 M_K$
   in Table~\ref{t:table1} (thereby also unbinding the isoscalar channel), 
   causes  $\sigma_{elas}$ to decrease by  $\sim 20\%$ from $106$ to
   $86$ mb while $\sigma_{a}$ remains unchanged.

   Calculations of the total elastic, reaction, and  $v$ times the annihilation cross 
   sections based on Eqs.~(\ref{e:physicalsig}) and (\ref{e:sigex})
   are shown in Fig.~\ref{f:fig5}. However, 
  as there are as yet no reported measurements of $K\bar K$ scattering with 
which to compare any of these cross sections, we  have attempted to use detailed
balance to derive the $\pi^+\pi^-\to K\bar K$ cross section, for 
which measurements do exist\cite{SDP73}. 
In this case the latter cross section can be exactly related to the
$I=0$ absorption cross section for $K\bar K\to\pi\pi$ defined by
   Eq.~(\ref{e:sigabsi}) as follows (see Eq.~(\ref{e:detail}))
\begin{eqnarray}
p^2 \sigma_a(\pi^+\pi^-\to K\bar K )
 = \frac{2}{3} k^2\sigma^{(0)}_a = \frac{2\pi}{3}(1-\eta^2_0)
\label{e:detbalance}
\end{eqnarray}
where $\eta_0 = |S_0(k)|$ is the $K\bar K$ inelasticity, and where $p$ and $k$ are the relative momenta of the pion or kaon pair  of
the same total CM energy. 

The two curves marked $(a)+(b)$ in Fig.~\ref{f:fig6} (the upper one includes 
Coulomb distortion\cite{LL58} via the factor 
$|f^{(-)}_c(k,d)|^{-2}\approx 2\pi\eta[exp(2\pi\eta)-1]^{-1}$  
taken from Eq.~(\ref{e:coulcorrection}) of the Appendix for $kd\sim \mu\alpha d\ll 1$)
show the inferred results   
 for $\sigma_a(\pi^+\pi^-\to K\bar K)$, where they are compared with  
 $\pi^+\pi^-$
scattering data\cite{SDP73,ADM77}, and the fit, curve $(c)$, to these data using a 
parametrized $\pi\pi$ scattering amplitude with parameters extracted from 
these experiments.  The extraction procedure is documented in detail
in\cite{SDP73}.  A coupled channel treatment of the $\pi\pi$--$K\bar K$ 
system leads to  similar results\cite{Lohse90,Sassen01,julich03}
(especially \cite{julich03}) shown by curves $(d)$ and $(e)$.

 Fig.~\ref{f:fig6} clearly shows that 
the strong enhancement of the cross section
near threshold is not adequately reproduced by the pure molecular picture. The same remark holds for the calculated $\pi\pi$ inelasticity,
discussed below, that is correspondingly somewhat too large near threshold ( Fig.~\ref{f:fig6a}).  

The pure molecular picture agrees reasonably well with the result obtained by that 
coupled  channel calculation of \cite{julich03} which { \it omits } the bare scalar meson state 
  $f_0^{\prime}$ (curve $(d)$). This numerical finding justifies the assumptions made
in order to obtain a simple analytical model.
In order to obtain a good agreement with the data, the introduction of an explicit bare scalar
meson state is necessary (curve $(e)$), however.

 The parametrized $\pi\pi$ scattering amplitudes as taken from 
 experiment\cite{SDP73} (curve (c)) assume the existence of a  pole  on the second 
  Riemann sheet just below the $K\bar K$ threshold at $[997\pm 6-(27\pm 8)i)]$ MeV in the $\pi\pi$ scattering 
  amplitude.
 The experimentally inferred value of this 
  pole lies reasonably close to the value of the 
   pole on the  second sheet for a decaying $f_0(980)$ bound 
  state given in Table \ref{t:table2}, that has been calculated from 
  the very different points of view
   presented here and elsewhere\cite{WS90}. 
  
 A further experimental check on these calculations is provided by 
comparing with the data \cite{BHyams1973}   
for the $\pi\pi$ isoscalar inelasticity parameter $\eta_0(\pi\pi)$. Since we have assumed that only coupling between 
the $\pi\pi$ and $K\bar K$ channels is important, this paramater
equals the $K\bar K$ inelasticity parameter $\eta_0(K\bar K)=|S_0(k)|$  that 
may be calculated directly from Eq.~(\ref{e:SI}).  The resulting prediction for
$\eta_0(\pi\pi)$ is given in Fig.~\ref{f:fig6a}.  
For comparison we also show the
behavior of $\eta_0(\pi\pi)$ that follows from a fully relativistic
treatment of the pion loop integral under Pauli--Villars regularization
with cutoff $\Lambda_{PV}=4$ GeV (the result is rather insensitive
to the actual choice of cutoff). There is no difference in the two 
calculations near
threshold. However, the relativistic version becomes more elastic than
the non--relativistic calculation with increasing CM energy.

 \section{Kaonium}

  Besides  the isoscalar $K\bar K$  bound system, the
 $K^+ K^-$ component of  this state can  bind to
  form  a  kaonium atom
  in the long range Coulomb potential.
  In the following we
  construct the Jost function that determines the  bound states and decay
  widths of kaonium, that may then be calculated without approximation.
  The hamiltonian
  now also includes the Coulomb field $V_c=\alpha Q\bar Q/r$,
   ($Q,\bar Q$ are  particle, antiparticle
  charge operators) that breaks isospin.  However, due to the very disparate length
  scales $\sim M^{-1}_K$ and $\sim (\alpha M_K)^{-1}$   over which the
 strong  and  much weaker Coulomb  interactions  operate, isospin breaking is
 essentially
 confined to that region of space $r\ge d$ outside the range of the
 strong potential.
 In this region the wave function  reads
 \begin{eqnarray}
  \frac{1}{r}\Big(\varphi_p|K^+K^->+\varphi_n|K^0\bar K^0>\Big),\quad r\ge d
 \label{e:extwf}
 \end{eqnarray}
 where $\varphi_p$ and $\varphi_n$ describe the $s$--wave radial motion of $K^+K^-$ and
 $K^0\bar K^0$ separately in the external region,
 \begin{eqnarray}
 &&(\frac{d^2}{dr^2}+k^2-M_KV_c)\varphi_p=0,\quad r>d
 \nonumber\\
 &&(\frac{d^2}{dr^2}+k^2)\varphi_n=0,\quad r>d
 \end{eqnarray}
Inside of this range we can ignore isospin breaking entirely
 due to the dominance of the strong isospin conserving interaction
 (no internal mixing
 approximation\cite{DR65,RHL69}) and work with states of good isospin.
  To obtain the Jost function we therefore
  need to join the linear combinations $-(\varphi_p\pm \varphi_n)/\sqrt 2$
  corresponding to $I=0$ and $1$
  in the external region $r\ge d$ onto the known  solutions
  for the complex potentials in Eq.~(\ref{e:newschr}) in the internal region $r\le d$.
  The latter are
   given by  $\varphi( k,r)$ of Eq.~(\ref{e:reg}) with the appropriate
   values of
   $a$,  $b$ and $\zeta$
   taken from Table~\ref{t:table1}
   for each isospin channel.

   For $r>d$ the solution describing the scattering of a $K^+K^-$ pair
    is given by the standard expression
   \begin{eqnarray}
   \varphi_p(r)= f^{(-)}_c(k,r)-S(k)e^{2i\sigma} f^{(+)}_c(k,r)
   \label{e:smatrix}
   \end{eqnarray}
    The $f^{(-)}_c( k,r)$ and $f^{(+)}_c(k,r)=f^{*(-)}_c(k,r)$ represent incoming (outgoing) waves in the Coulomb
    field, $S(k)$ is the scattering matrix of non--Coulombic origin,
     $\sigma=arg\Gamma(1+i\eta)$  the Coulomb phase shift and
   $i\eta=\mu\alpha/ik$ the Coulomb parameter for attractive
   interactions; $\mu=M_K/2$ is the reduced mass.
    The $f^{(\pm)}_c(k,r)$  are  proportional
   to  Whittaker functions. One has\cite{RGN66}
   \begin{eqnarray}
   && f^{(-)}_c(k,r)=e^{\pi \eta/2} W_{i\eta,\frac{1}{2}}(2ikr)
    \nonumber\\
   &&\approx
    \frac{e^{\pi \eta/2}}{\Gamma(1-i\eta)}
    \Big\{1-ikr-2\mu\alpha r[\ln 2ikr+
    \psi(1-i\eta)+2\gamma-1]+\cdots\Big\},\quad kr\to 0
   \label{e:fc}
   \end{eqnarray}
   where the small $kr$ behavior of $f^{(-)}_c(k,r)$ involves\cite{GR80}  the
   standard $\psi$--function, and $\gamma=0.57721$ is the Euler constant.

      The $S$--matrix in Eq.~(\ref{e:smatrix}) is  determined by
      the logarithmic derivative $\zeta_p=\varphi^\prime_p/\varphi_p$ of $\varphi_p$
      at
      $r=d$. In detail,
      \begin{eqnarray}
      S(k)=e^{-2i(\sigma-\rho)} \Big(\frac{\zeta_p-\zeta_c}{\zeta_p-\zeta^*_c}\Big)
      \label{e:coulsmatrix}
      \end{eqnarray}
      where $\zeta_c=f^{\prime(-)}_c/f^{(-)}_c$ is the logarithmic derivative
      of the incoming Coulomb wave at $r=d$, and $\rho =\arg f^{(-)}_c(k,d)$.
       Thus the Jost function  describing
      kaonium is proportional to $\zeta_p-\zeta_c$. To find it\cite{RHL69}
      we determine the
      value of $\zeta_p$ by matching
      $\varphi_p\pm\varphi_n$ onto the internal solutions $\varphi_0$ and
    $\varphi_1$ of good isospin,
    where $\varphi_n\sim exp(ikr)$ is a
    pure outgoing  wave\footnote{This channel is actually closed and decays exponentially
    like $\sim exp(-\sqrt{2M_{K^0} \Delta}\: r)$  when isospin breaking due to the
    $K^0$--$K^\pm$ mass difference $\Delta\sim 4$MeV is not ignored.
    This is an inessential approximation in the presence of the dominating isoscalar
    channel. Keeping $\Delta$ changes the
     eigenvalues of Eq.~(\ref{e:lroots}) by at most $\sim 10\%$, see below
     Eq.~(\ref{e:eigenrevised}).}.
    Due the large difference
   in space behavior of the strong $(br\gg 1)$ and Coulombic $(kr\ll 1)$ wave functions
    near $r\sim d$ for momenta $k\sim \mu\alpha$ of interest for kaonium,
   the $\varphi(k,r)$ have already reached their asymptotic behavior,
     while the $f^{(\pm)}_c(k,r)$  are still given by
   their
   small $kr$ approximation. The matching of value and derivative at $r=d$
   is therefore particularly simple to perform. One finds
   \begin{eqnarray}
      \frac{1}{\zeta_p}= \frac{1}{2}\Big(\frac{1}{\zeta_0}+\frac{1}{\zeta_1}\Big)
      \label{e:zetap}
   \end{eqnarray}
 where the logarithmic derivatives $\zeta_I=\varphi_I^\prime/\varphi_I$
 of the internal wave functions carrying good isospin $I$
  are
\begin{eqnarray}
\zeta_I= ik\Big(\frac{f(k)e^{ikd}+f(-k)e^{-ikd}}{f(k)e^{ikd}
-f(-k)e^{-ikd}}\Big)_I
\label{e:zetaI}
  \end{eqnarray}

  The eigenvalue condition $\zeta_p=\zeta_c$ that determines complex energies
  of kaonium then reads
\begin{eqnarray}
1-\frac{1}{2}\zeta_c(\frac{1}{\zeta_0}+\frac{1}{\zeta_1}) =0
\label{e:eigenvalues}
\end{eqnarray}

 This equation contains no free
parameters. The  Jost functions $f(k)$ in
 the $\zeta_I$ all have pre--determined
 constants  for each isospin channel as given in
  Table~\ref{t:table1}, and come from Eq.~(\ref{e:revjost})
 if annihilation is included, or Eq.~(\ref{e:jostfcn}) if it is not.

 It is useful to express the roots of Eq.~(\ref{e:eigenvalues}) in the form
 $k_\lambda=-i\lambda\mu\alpha$ since then $\lambda^{-1}\to 1,2,3,...$ for a pure Coulomb field,
 and also to recognise that the logarithmic derivatives $\zeta_I$ can
 be replaced by
 their  zero momentum values
 with impunity because
 $k$ is always ${\cal O}(\mu\alpha)$ for kaonium.  Then
 \begin{eqnarray}
  \frac{1}{\zeta_I}\approx-(a^{eff}_I-d)+O(1/\zeta^2)
 \label{e:zerolndrv}
 \end{eqnarray}
 in terms of the complex scattering length of Eq.~(\ref{e:cmplxln}).

  Using these forms for the $\zeta_I$, the eigenvalue condition
  (\ref{e:eigenvalues}) can be reformulated in terms of the complex scattering
  length  $a^{eff}_p$ for $K^+K^-$ scattering as
   \begin{eqnarray}
  a^{eff}_p=\frac{1}{2}(a^{eff}_0+a^{eff}_1)=\Big(1-\frac{1}{\zeta_c d}\Big)d
 \label{e:second}
 \end{eqnarray}
 This leads to an  eigenvalue equation
 of the  Kudryavtsev--Popov (KP)  type\cite{KP79},
 \begin{eqnarray}
 1/a^{eff}_p=
  2\mu\alpha\Big[\psi(1-1/\lambda)+\frac{1}{2}\lambda
  +\ln (2\lambda\mu\alpha d)+2\gamma\Big]
\label{e:lroots}
\end{eqnarray}
The logarithmic derivative $\zeta_c(k)$ of the Coulomb wave
function has been calculated at $k_\lambda=-i\lambda\mu\alpha$ from
Eq.~(\ref{e:fc}).

The  resulting  complex energy spectrum for kaonium is then given by
\begin{eqnarray}
  E_\lambda-\frac{i}{2}\Gamma_\lambda=-\frac{1}{4}\lambda^2 M_K\alpha^2
\label{e:kaoniumspec}
\end{eqnarray}
The solutions of the KP equation are dominated by the $\psi$--function
on the right hand side that has poles at $\lambda=1/n$, for integer $n$.
These poles just reproduce the pure Coulomb spectrum of kaonium.  The strong interaction
changes this behavior by moving $\lambda$ off these poles to
$\lambda=(n+\delta)^{-1}$, where $\delta$ is complex and not necessarily small.
In Table~\ref{t:table4} we list a series
 of exact numerical eigenvalues $\lambda$ of the KP equation, and
 the associated complex energy spectrum  for kaonium. These computations use
 the  value  
 \begin{eqnarray}
 a^{eff}_p=(1.940-1.199i)M^{-1}_K =(0.77-0.48i) {\rm fm}
 \label{e:apeff}
 \end{eqnarray}
 for the complex scattering length  calculated from Eq.~(\ref{e:cmplxln}) plus the data 
 given in Table~\ref{t:table3}. Notice that the $a_0(980)$ admixture introduced via the isovector 
   scattering length
   $a_1$ only contributes to the real part of the $K^+K^-$ effective 
   scattering length that enters into the calculation of the level shifts 
   and decay widths 
   of kaonium. The imaginary part is entirely determined by the isoscalar piece of the
   strong interactions. However, due to the non--linear nature of the eigenvalue 
   problem, the isovector contribution nevertheless influences
   both the real and imaginary parts of the resulting energy spectrum. Only 
   in the perturbative limit (to be described  
   below)  does the $a_0(980)$ mode contribute solely to the level shifts.

 For comparison we have also used the isospin zero and one scattering lengths
 $\hat a^{eff}_0= (0.86-1.82i)$ fm and  $\hat a^{eff}_1= (0.06-0.77i)$ fm, 
 based on a recent investigation\cite{julich03} into
 the effects of a high lying scalar isoscalar quark antiquark
 configuration (which we refer to as $f_0^\prime$) mixed into the
 $K\bar K$ molecular ground state. Here the effect of the
 $\pi\eta$-channel is included too.
 The  $K^+K^-$ scattering length then becomes
 \begin{eqnarray}
  \hat a^{eff}_p=(1.156-3.268i)M^{-1}_K =(0.46-1.30i) {\rm fm}   
 \label{e:felixapeff}
 \end{eqnarray}
 instead of Eq.~(\ref{e:apeff}). 
 
The predictions for kaonium 
 are given in Tables~\ref{t:table4} and \ref{t:table5}, while Table~\ref{t:table6} 
 summarizes the decrease in lifetime of the ground state as the complexity of
 the $K^+K^-$ interaction is increased.  In particular,
Table VI shows that the lifetime of kaonium is reduced further by about 
a factor of three once annihilation in the isovector channel is introduced
via the $\pi\eta$ decay of the $a_0(980)$. These results are qualitatively
 similar to those given by other approaches to the problem of kaonium
 formation and decay\cite{SWycech}.
One notes from Tables~\ref{t:table4} and \ref{t:table5} that 
Re$\lambda< 1/n$, corresponding to a  positive shift
 of the Coulomb level when the strong interaction supports a bound
 state (the sign of the real part of the scattering length determines the
 sign of the level shift).
In addition to causing repulsive shifts of the Coulomb levels
the  strong field  introduces a decay width due to $K^+K^-\to 2\pi$
annihilation with a lifetime of $\sim 10^{-18}$ sec
(the two--photon annihilation
 lifetime at ${\cal O}(M^{-1}_K\alpha^{-5}) \sim 10^{-13}$ sec,
 or one in $10^5$ decays, is too long to compete).
It also introduces  an extra node
into  all the eigenfunctions of  kaonium. This means that the kaonium
levels represent
 {\it excited} states in the combined Coulomb
plus strong
fields of the  $K\bar K$ system.  As will be discussed in more detail below, 
the occurrence
of this node  is responsible for making the kaonium lifetime 
  two orders of magnitude shorter than the $\sim 10^{-16}$ sec 
 it would have been  had the $K^+K^-$ pair 
 only been bound by a Coulomb 
field\footnote{The predicted lifetime of kaonium  is  three orders of magnitude
shorter than that for the  pionium $\to 2\pi^0$ decay\cite{GLRG01}.
Apart from specific details of the annihilation interaction for pionium,
its decay rate
 is slowed down by the relatively
small mass difference $M_{\pi^+}-M_{\pi^0} \sim 5$ MeV  that drives the decay
as opposed to $M_K-M_\pi\sim 356$  MeV for kaonium. Also, the $\pi\pi$ system
has no bound state analogous to $K\bar K$.}.

Let us briefly indicate how these features come about.
The appearance  of the additional  radial node is required by the fact that
the $1s$ state of relative motion
has been usurped by the $K\bar K$ ground state in the strong potential. So
 the ground and excited levels of kaonium have to have a $2s,3s...$
character with one additional
node. This node is   always  positioned in  the  close proximity
of the $K^+K^-$
scattering length, independent of the excitation energy. It is real or virtual
depending on whether the strong interaction supports a
real ($a_p>0$) or a virtual ($a_p<0$) bound state.
 This comes about  as follows. The
 kaonium levels are all weakly bound at $k\sim \mu\alpha$
  so that
their radial wave functions in the inner region
 all become nearly identical
with the  zero energy wave functions of $K\bar K$ in the strong potential alone. These
functions [see Eq.~(\ref{e:limitwf})]
all have a node  at $r=a_p$ in the effective range approximation.  The remaining
nodes are  supplied by the Whittaker function
 $W_{1/\lambda,\frac{1}{2}}(2\lambda \mu\alpha r)$ in the external region,
 $r\ge d$.
 They are almost identical with those of a pure Coulomb field bound state
 wave function where $\lambda=1/n$.

The eigenfunction for the lowest  state of kaonium 
 is illustrated in Fig.~\ref{f:fig7} (with absorption suppressed).
It is clear from this figure that the actual choice of the cutoff $d$
is immaterial if taken anywhere in the region where the wave function behaves
linearly.
A closed form for the bound state eigenfunctions $\varphi_{1/\lambda}$ of kaonium labelled by the
quantum number $\lambda^{-1}$ is given in Eq.~(\ref{e:kaoniumwf})
in the Appendix. From that one calculates
 that the
gradient
at the origin is enhanced by the factor $b/a$ in both isospin channels
over that of the pure Coulomb state. 
  Physically this comes about  because the strong potentials
 draw the $K^+K^-$ pair much closer together,
 thereby
 increasing the probability density of finding them
at the origin by
 $(b^2/a^2)_I$ (notice $b>|a|$ from Eq.~(\ref{e:ab})) in each isospin
 component, where they annihilate.  In the isoscalar channel, which is strongly attractive,
 the effect is especially marked, $(b^2/a^2)_0\sim 60$. On the other hand the
 isovector channel only supports a virtual state. Hence the concentration of
 probability density at the origin in this channel is much weaker, 
 $(b^2/a^2)_1\sim 3$. 

   The effect of this factor in the isoscalar channel is quite dramatic. It 
 increases the decay 
 width of kaonium by two orders of magnitude (with a corresponding 
 decrease in lifetime) 
 over what it would have been in the 
 absence of the strong potentials $V_I$. One can illustrate this
  qualitatively by appealing to the
 perturbative solution\cite{KP79,Deser54} of
 the KP equation for the complex energy shift.
 For the ground state this reads
  \begin{eqnarray}
 \Delta E-\frac{i}{2}\Gamma\approx \frac{1}{2}M^2_K\alpha^3a^{eff}_p
= \frac{1}{2}M^2_K\alpha^3a_p-\frac{1}{2}M^2_K\alpha^3\xi^{-1}
\big[ \frac{1}{2}(\frac{b^2}{a^2})_0\big]
 \label{e:approxKP}
 \end{eqnarray}
 that equals either $(0.187-0.116i)$ keV, or  $(0.187-0.164i)$ keV
 depending on whether the value (79) or its low 
  absorption version $\xi\gg 1$ taken from
 Eq.~(\ref{e:cmplxln}),
 \begin{eqnarray}
 a^{eff}_p\approx \frac{(a_0+a_1)}{2}-\frac{i}{2}(\frac{b^2}{a^2\xi})_0 
 =(1.940-1.703)M^{-1}_K
 \end{eqnarray}
 is used for the  effective scattering length.
 The $a_I$ are the bare scattering lengths of 
 Eq.~(\ref{e:bargar}) as before.  The last approximation places the role of 
 the isoscalar
 probability density admixture $(b^2/a^2)_0$  in the kaonium ground state
 explicity in evidence. 
 
 Neither approximation is 
 very accurate in comparison with the exact solution of $(0.181-0.107i)$ keV, 
 but can be used to illustrate the effect of 
 strong interactions
 decreasing the lifetime of kaonium against what it would have been
  in a pure Coulomb field.
One sees this as follows. The decay width in Eq.~(\ref{e:approxKP}) is directly related to
 the ``annihilation at rest'' cross section 
 \begin{eqnarray}
  \sigma_a\approx -\frac{4\pi}{k}Im(a^{eff}_p)=\frac{4\pi}{k}
  \frac{1}{2}\big[\frac{b^2\xi}{a^2\xi^2+(b^2-a^2)^2}\big]_0
 \approx \frac{4\pi}{k\xi}\big[ \frac{1}{2}(\frac{b^2}{a^2})_0\big] 
 \label{e:xpn}
 \end{eqnarray}
 as $\Gamma=v\phi^2(0)\sigma_a$, upon multiplication by the flux factor
 $v\times \phi^2(0)$ where $\phi^2(0)=M^3_K\alpha^3/8\pi$  gives the 
 probability per unit volume for finding the $K^+K^-$ pair at the origin
 when bound by the Coulomb field alone; $v=k/\mu$ is their relative velocity.
 On the other hand when
  the strong potential is absent, the scattering
   lengths $a^{eff}_0\to -i/\xi$ and $a^{eff}_1\to 0$  are pure imaginary and zero
   respectively, so that $a^{eff}_p\to -i/2\xi =-0.028iM^{-1}_K$, and the
   corresponding free annihilation cross section  becomes
   \begin{eqnarray}
   \sigma_{a,free}=\frac{4\pi}{k\xi}(\frac{1}{2}) 
   \approx \sigma_a/\theta,\quad \theta\sim 42\;{\rm to}\;60 
   \end{eqnarray}
  if we use the unexpanded or expanded version of $\sigma_a$ in 
  Eq.~(\ref{e:xpn}). The factor $\theta$ reflects the modification of the 
  kaonium ground state due to the strong interactions.
  
  In this case there is no energy shift at all, and the
   decay half--width is $\frac{1}{2}\Gamma_{free}= 0.0028$ keV, leading to a 
   lifetime of $\tau_{free}=1.2\times 10^{-16}$ sec.
 So the $K^+K^-\to 2\pi$ strong potential assisted annihilation lifetime 
 drops by roughly the factor $\theta$ from $1.2\times 10^{-16}$ sec that it would 
 have been in a pure Coulomb field, to $3.2\times 10^{-18}$ sec as given by 
 the exact
 solution of the KP equation. 
 Table~\ref{t:table6} summarizes the change in lifetime       
 of the kaonium ground state as the
  complexity of the $K^+K^-$ interaction
 changes from assuming free annihilation at rest in a Coulomb field, to
 including one--boson exchange, to adding $f^\prime_0$ mixing, for determining 
 the scattering length.

 We note in passing from the center
 expression for $\sigma_a$ in Eq.~(\ref{e:xpn}) that if the  
  bound state occurs at zero energy ($a=0$), the corresponding enhancement
 of the annihilation width is maximal.
 Also, while generally true of the exact
 solutions of the KP equation, the perturbative solution given in
 Eq.~(\ref{e:approxKP}) makes it evident
 that  the sign of the level shift goes along with the sign of the
 scattering length, which is positive for bound and negative for unbound
 or virtual states.  

 We close this section with a comment on to what extent the results for kaonium
 might be considered as ''parameter--free''. Apart from specifying a 
  Lagrangian  with the attendant assumption of $SU(3)$ symmetry that fixes 
  the various coupling constants in terms of $g_{\rho\pi\pi}$, that we 
  in turn fixed from
  the KSRF relation, the only arbitrary parameter in these calculations
  was the form factor cutoff $\Lambda$ introduced in Section IIIA. The choice $\Lambda=
  4$ GeV, that was also used in Ref.~\cite{Lohse90}, led to  calculated scattering 
  lengths and effective ranges for the $K\bar K$ system, that were then
  duplicated by constructing  an equivalent Bargmann potential that depended
  on two parameters $a$ and $b$. This in turn gave predictions for the 
  mass and width of a bound $K\bar K$ molecular state in reasonable 
  accord with the known experimental  mass and width of the  $f_0(980)$ scalar meson.  
  However, as already demonstrated in Table~\ref{t:table2}, one can turn the argument
  around without materially affecting any of our conclusions for kaonium, 
  by asserting that the Bargmann parameters $a$ and $b$ are actually 
  pre--determined from experiment in the
  strong interaction sector, thus leaving the kaonium sector without any
  free parameters at all.  
   
   Similar remarks hold true for the calculated  cross sections shown 
   in Figs.~\ref{f:fig5} and \ref{f:fig6}. Once $a$ and $b$ have been fixed from
   experiment or otherwise, no further adjustable parameters enter into these
   calculations. The actual values of $a$ and $b$
  as determined by any of the five alternatives shown in Table~\ref{t:table2}
  do not in fact differ very much.
 
  \section{Discussion}

 In this investigation we have shown that  one meson exchange
 potentials derived from a standard $SU(3)_V\times SU(3)_A$
  Lagrangian in the non--relativistic limit are sufficient  to bind $K\bar K$ into a
 kaonic molecule with a mass and decay width that closely match the experimental
 values of the $f_0(980)$ isoscalar meson.  We have also shown that the same potentials,
 in combination with detailed balance, lead to a prediction of 
 the $\pi^+\pi^-\to K\bar K$ cross section that is
  too low especially near threshold.  

 We have further demonstrated
 that there are specific, experimentally observable effects
 in the
 energy level shifts and decay properties of the  kaonium spectrum,
 in the presence of a $K\bar K$
  molecular  ground state.
 The existence of such a
  bound  $K\bar K$  pair would be reflected in positive energy level
  shifts in the kaonium Coulomb spectrum (due to level repulsion), plus
  enhanced $K^+K^-\to 2\pi$ decay
  widths. On the other hand,  unbound or virtually bound
  (and therefore weakly correlated) $K\bar K$
  pairs  produce negative energy level shifts and a
  weak enhancement  of the decay widths. Since the sign of the level shifts and the
  enhancement of the decay widths go together with the presence or absence of
  a strongly correlated $K\bar K$ pair in the ground state, measurements
  of the level shifts and decay widths in kaonium could provide another  
  probe for investigating to what extent the $f_0(980)$ meson
  ground state has a $K\bar K$ molecular component. 
  From the experimental point of view it 
  is important
  to stress that, as for protonium\cite{KP79} which has a very similar 
  decay width, the $1s-2s$ level
  spacing of $\sim 5.1$keV in kaonium is many times the estimated decay 
  widths $\lesssim 200$ eV, so that these levels remain distinguishable in the 
  presence of the $2\pi$ decay channel.
           
  Notwithstanding the results reported in \cite{Barnes1985}, we again stress
  the conflict that exists between the pure molecular 
  description of the  $f_0(980)$ ground state and the calculated $2\gamma$
  decay width. In \cite{Barnes1985} a binding energy of $10$ MeV for the kaonium molecule was 
  assumed, that leads to a width of $\sim 0.6$ keV. However, most 
  recent data show a resonance energy that corresponds to a pole in the 
  complex plane, which translates into a binding energy  of $18$ to $20$ MeV
  for the molecular potential. This increase in binding energy enhances the 
  value of the spatial wave function at the origin by a factor of three, see
  Fig.~\ref{f:fig8}. One way of understanding the origin of this discrepancy is
  to generalise the molecular picture of the $f_0(980)$ to a superposition 
  of a $K\bar K$ component and a hard component. In the J\"ulich model
  \cite{Lohse90} a $f^\prime_0$ state has been considered in addition to the
  pure molecule in order to reproduce the large inelasticities of the $\pi\pi$ phase 
  shifts in the vicinity of the $K\bar K$ threshold. The $f^\prime_0$ state may be 
  interpreted as either a $q\bar q$ or as a
  $q^2\bar q^2$ configuration. As pointed out by Achasov\cite{achasov82} and 
  confirmed by current algebra\cite{RHLRT2002}, the large  decay width of the former
  into two pions may disqualify a $q\bar q$ interpretation of the additional component
  $f^\prime_0$.  
   The effect of the $f^\prime_0$ component in the $f_0(980)$ meson introduced in the
   J\"ulich model is to reduce the lifetime of kaonium from $3.2\times 10^{-18}$
   sec to $1.1\times 10^{-18}$ sec.                           
  The presently available data concerning the 
  inelasticities of the $\pi\pi$ phase shifts, and in particular the 
  $\pi^+\pi^-\to K\bar K$ cross sections, do not allow one to disentangle the
  components of the $f_0(980)$ structure in a model independent way.  High resolution data for two kaon
  production near to the threshold will be undertaken at COSY in the 
  near future.

\section{Acknowledgements}
 One of us (RHL) would like to thank Prof. Speth of the Institut f\"ur Kernphysik
 Theory Division, Forschungszentrum, J\"ulich, for the exceptional hospitality of the
 institute, and for financial support. It is a pleasure to thank
 M. B\"uscher, D. Gotta, C. Hanhart and A. E.Kudryavtsev for useful discussions.

\newpage
\appendix*
\section{}
\label{a:A}
\subsection{Isospin phase conventions}
In order to construct two--meson states of good isospin using conventional
Clebsch--Gordan coefficients the following phase conventions are useful.

If one chooses\cite{Petersen71} as basis states $|\pi^\pm>=\mp|\pi,\pm1>$, and
$|\pi^0>=|\pi,0>$ for charged and neutral pions, the  two--pion state
 of good isospin $|I=0,I_3=0>$ is, for example,
\begin{eqnarray}
|(\pi\pi)_0>=-\sqrt{\frac{1}{3}}\Big[|\pi^+(1)\pi^-(2)> +|\pi^-(1)\pi^
+(2)>+|\pi^0(1)\pi^0(2)>\Big]
\label{e:pipi}
\end{eqnarray}
Since the pions are identical bosons in the isospin basis, one has to
include a
space part of the two--pion wave function that is symmetric in particle
label
interchange to accompany $|I=0,I_3=0>$. The normalization of the
symmetrized product of the two--pion space wave function introduces
an additional factor $n_s=1/\sqrt 2$.

For the kaon  and anti--kaon  isospin doublet and anti--doublet
the analogous
convention is, in matrix form,
\[K=\left(\begin{array}{c}
K^+\\
K^0\\
\end{array}\right ),\quad \bar K=
 \left(\begin{array}{c}
 \bar K^0\\
 -K^-\\
  \end{array}\right )
 \]
Consequently the   isoscalar and isovector combinations $|I=0,I_3=0>$
and $|I=1,I_3=0>$ are
\begin{eqnarray}
&&|(K\bar K)_{0}>=-\frac{1}{\sqrt 2}[|K^+(1)K^-(2)>+|K^0(1)\bar K^0(2)>]
\nonumber\\
&&|(K\bar K)_1>=-\frac{1}{\sqrt 2}[|K^+(1)K^-(2)>-|K^0(1)\bar K^0(2)>]
\label{e:KK}
\end{eqnarray}
These relations lead immediately to the values of the isospin coefficients $C_I$
and $n_s$ quoted below Eq.~(\ref{e:gammahat}) in the text.

\subsection{Wave functions for the Bargmann  potentials}
We briefly summarize some results of Bargmann
for constructing phase equivalent potentials. As shown in\cite{Bargmann49}
the assumed form Eq.~(\ref{e:jostfcn}) for the Jost function gives rise
to a family of potentials that lead to this form for $f(k)$: the simplest one
of this family is given by Eq.~(\ref{e:bargpot}).
Let $\psi(r)=r^{-1}\varphi(r)$, and replace the OBE potentials by this
 form for $V_I$. Then
\begin{eqnarray}
\varphi^{\prime\prime}+(k^2-M_K V_I)\varphi=0
\label{e:swaveschr}
\end{eqnarray}
for $K\bar K$ $s$--wave  scattering.
The two linearly independent solutions $f(\pm k,r)$ for this equation
 that are irregular at the origin are known in closed form by 
 construction\cite{Bargmann49,Newton60}
\begin{eqnarray}
 f(k,r)&&=e^{-ikr}\frac{k-ib[b\sinh(br)+a\cosh(br)][b\cosh(br)+a\sinh(br)]^{-1}}{k-ib}
\nonumber
\\&&\to e^{-ikr},\quad r\to\infty
\label{e:irreg}
\end{eqnarray}
A general solution of Eq.~(\ref{e:swaveschr})
can be written as a linear combination of the $f(\pm k,r)$,
\begin{eqnarray}
\varphi(k,r)=\frac{i}{2k}[f(-k)f(k,r)-f(k)f(-k,r)]
\label{e:reg}
\end{eqnarray}
 where the coefficient $f(k)$ is the Jost function. It can be  shown quite
 generally\cite{Newton60} that  $f(k)$ is determined by the Wronskian
 of $f(k,r)$ and $\varphi(k,r)$:
 \begin{eqnarray}
 f(k)=W[f(k,r),\varphi(k,r)] =[f\varphi^\prime-f^\prime\varphi]
 \label{e:defjost}
 \end{eqnarray}
 Since $W$ is independent of $r$ this can be evaluated at any convenient
 point, in particular at the origin.
 The value of $f(k)$  is determined  by how
 $\varphi$ behaves at the origin. If $\varphi\sim r$, then
  $f(k)=f(k,0)=(k-ia)/(k-ib)$ as in
 Eq.~(\ref{e:jostfcn}).  This is the usual case.
 On the other hand if
   $\varphi$ does not vanish at the origin,
   but instead possess a finite logarithmic derivative there,
  then $f(k)$ is given by Eq.~(\ref{e:revjost}).

 In the first case  one obtains the usual scattering solution of
 Eq.~(\ref{e:swaveschr}) that vanishes at the origin as
 \begin{eqnarray}
 \varphi(k,r)&&=\frac{\sin kr}{k}+\frac{b^2-a^2}{b^2+k^2}
 \frac{1}{k}\frac{k\sinh(br)\cos(kr)-b\cosh(br)
 \sin(kr)}{a\sinh(br)+b\cosh(br)}
 \label{e:bargscatt}
 \end{eqnarray}
If  $a=-\kappa$ is negative, then $f(k)$ has a zero in the lower half of the
 complex $k$-plane, corresponding to a
 bound state with binding energy $\varepsilon=\kappa^2/M_K$. The associated
 bound state wave
 function is found by setting $a=-\kappa$ and $k=-i\kappa$ in
 Eq.~(\ref{e:bargscatt}) to find
 \begin{eqnarray}
 \varphi(-i\kappa,r)&&=\frac{\sinh(br)}{b\cosh(br)-\kappa\sinh(br)} e^{-\kappa r}
 \end{eqnarray}
 The normalization of this bound state is also available from a knowledge of the
 Jost function:
 \begin{eqnarray}
 \int_0^\infty dr\varphi^2(-i\kappa,r)=\frac{i}{4\kappa^2}f(i\kappa)\frac{\partial
 f(k)}{\partial k}_{k\to -i\kappa}
 = \frac{i}{4\kappa^2}f(i\kappa)\dot f(-i\kappa)=\frac{1}{2\kappa(b^2-\kappa^2)}
 \label{:ebargwfu}
\end{eqnarray}
since  $f(k)=(k+i\kappa)/(k-ib)$.  The  normalized bound
state  is then given by
\begin{eqnarray}
\psi(r)=(\sqrt{4\pi}r)^{-1}
\sqrt{2\kappa(b^2-\kappa^2)} \frac{\sinh(br)}{b\cosh(br)-\kappa\sinh(br)}
e^{-\kappa r}
\label{e:bargu}
\end{eqnarray}

In Fig.~\ref{f:fig8} we compare the radial eigenfunction as obtained
by numerically solving for the $K\bar{K}$ isoscalar bound state in the
one boson exchange potential of Eq.~(\ref{e:OBE}), with the analytical
solution given by Eq.~(\ref{e:bargu}) for the phase equivalent
Bargmann potential. Except in the vicinity of the origin, where the
numerical solution is underestimated by a factor $\sim 1.2$, the two
wave functions are essentially identical. Also shown is the wave
function at $10\textrm{ MeV}$ binding in a gaussian potential that was
used in \cite{Barnes1985} to estimate $\psi(0)$.

Finally in Fig.~\ref{f:fig9} we illustrate the variation of the
spatial wave function $\psi(0)$ of Eq.~(\ref{e:bargu}) as a function
of its binding energy. This plot was obtained by using the calculated
variation of $a_0$, $r_0$ and $\varepsilon$ with cutoff as shown in
Fig.~\ref{f:fig2} for the one boson exchange potential to obtain the
parameters $a=-\kappa$ and $b$ at each binding energy.

The inclusion of the annihilation potential changes the solution for the
radial eigenfunction from $\varphi(-i\kappa,r)$ to the complex solution
\begin{eqnarray}
\tilde\varphi(k_r,r)=f(k_r,r)/f(k_r,0)\quad 
\end{eqnarray}
where $k_r = (0.1254-0.1972i) M_K$ is the  root of the Jost function in the
presence of annihilation as given by Eq.~(\ref{e:exactroot}).  
One readily verifies that $\tilde\varphi(k_r,0)=1$ and  
$\tilde\varphi^\prime(k_r,0)=\zeta=-i\xi$ as required. A comparative plot 
of the normalized versions of the
moduli of $\tilde\varphi(k_r,r)$ and  
 $\varphi(-i\kappa,r)$
 is given in Fig.~\ref{f:fig7a}.  Since the derivative of
$\tilde\varphi(k_r,r)$ at the origin is  purely imaginary, its modulus
 has a zero derivative there. This is clearly visible in the
inset in Fig.~\ref{f:fig7a}, that also shows how $|\tilde\varphi(k_r,r)|$ 
``heals'' within $0.1$ fm away from the origin from the effects of the annihilation potential 
to rejoin the real solution  $\varphi(-i\kappa,r)$. 
Either eigenfunction leads to
a rms molecular radius of $\sqrt{<r^2>}\approx 1.62$ fm, 
(i.e. about two proton charge 
radii). 

The momentum content of the normalized spatial wave functions
$\psi(r)=(\sqrt{4\pi}r)^{-1} \varphi(-i\kappa,r)$ and 
$\tilde\psi(r)=(\sqrt{4\pi}r)^{-1} \tilde\varphi(k_r,r)$ is given by their
fourier transforms $\psi(p)$ and $\tilde\psi(p)$.  These are shown in Fig.~\ref{f:fig7b}
for relative meson momenta up to $1$ GeV.
As expected from Fig.~\ref{f:fig7a}, the effect of the annihilation potential is to 
redistribute the momentum to higher values, while at the same time leaving 
the characteristic value of $<p^2>$ at
$\sim 0.08$ GeV$^2$ for the relative meson momentum squared in the molecule.

 \subsection{Matching of isospin amplitudes}
 The  wave functions of  kaonium  in the inner, or isospin conserving
region  are given by  Eq.~(\ref{e:reg}).
These solutions either vanish at the origin, or possess
a given logarithmic there, depending on the form of $f(k)$ that is chosen.
Matching either form  onto the linear
combinations $\varphi\pm \varphi_n$ of the outside solutions at $r=d$,
 one finds the consistency
condition
\[\left(\begin{array}{cc}
\zeta_0-\zeta_p&\zeta_0-\zeta_n\\
\zeta_1-\zeta_p&-\zeta_1+\zeta_n\\
 \end{array}\right )
 \left(\begin{array}{c}
 \varphi_p\\
 \varphi_n\\
  \end{array}\right ) =0
 \]
 where $\zeta_n=\varphi_n^\prime/\varphi_n$. The other logarithmic derivatives have been
 defined in the main text.   At momenta $k\sim \mu\alpha$ of relevance for
 kaonium,
 $\zeta_n=ik$ is much smaller than any of
  these so we simply drop it and find
 \begin{eqnarray}
 \frac{1}{\zeta_p}\approx \frac{1}{2}\big(\frac{1}{\zeta_0}
 +\frac{1}{\zeta_1}\big);\quad
 \varphi_n/\varphi_p\approx -1+\zeta_p/\zeta_0=
 1-\zeta_p/\zeta_1
 \label{e:consistency}
 \end{eqnarray}
The eigenvalue condition $\zeta_p=\zeta_c$
determines the values of $k=-i\kappa$, which are pure imaginary in the absence
of absorption, for the bound states of kaonium, and at the same time
removes the outgoing wave piece of $\varphi_p$ from Eq.~(\ref{e:smatrix}).
Then the matching conditions show that the kaonium eigenfunctions
 that join smoothly onto the pure decaying state $f^{(-)}_c(-i\kappa r)\sim
W_{1/\lambda,\frac{1}{2}}(2\kappa r)$ outside the strong potential are given by a
linear combination of isospin zero and one internal functions as
\begin{eqnarray}
 \varphi_{1/\lambda}(r)&&=\frac{1}{2}\zeta_c f_c
 \Big[\frac{1}{\zeta_0}\frac{\varphi_0(-i\kappa, r)}{\varphi_0(-i\kappa, d)}+
 \frac{1}{\zeta_1}\frac{\varphi_1(-i\kappa, r)}{\varphi_1(-i\kappa, d)}\Big],
 \quad r\le d
\nonumber\\
&&=f_c(-i\kappa, r),\quad r\ge d
\label{e:kaoniumwf}
\end{eqnarray}
 where $\zeta_c=\zeta_c(-i\kappa)$ and $f_c=f^{(-)}_c(-i\kappa,d)$. We have labelled these eigenfunctions with the ``quantum number''
$\lambda^{-1}=\mu\alpha/\kappa$ that reverts to an integer for a pure Coulomb field.
One verifies immediately that the matching of value at $r=d$ is guaranteed by the
eigenvalue condition (\ref{e:eigenvalues}) or (\ref{e:consistency}),
while the derivatives match identically from the definition of the
logarithmic derivatives.

 These eigenfunctions  develop a common additional node
 at the  $K^+K^-$ scattering length, independent
 of the energy of the kaonium level. Consider the case where
 the scattering length is real. Then this happens because
  $\kappa\sim \mu\alpha \ll M_K$ is very small so that the behavior of the
 $\varphi_I(-i\kappa,r)$  differ but little
 from their zero energy behavior. From Eq.~(\ref{e:bargscatt}) this is
\begin{eqnarray}
\varphi(0,r)&&=\frac{b-a}{b^2}\Big\{\frac{(b+a)\sinh(br)}{b\cosh(br)+a\sinh(br)}
+r\frac{ba}{b-a}\frac{b\sinh(br)+a\cosh(br)}{b\cosh(br)+a\sinh(br)}\Big\}
\nonumber\\
&&\sim r,\quad br\ll 1
\nonumber\\
&&\sim \frac{a}{b}(r-a_I),\quad br\gg 1
\label{e:zerowf}
\end{eqnarray}
where $a$ and $b$ refer to channel $I$.
Inserting this information into Eq.~(\ref{e:kaoniumwf}), one finds that
\begin{eqnarray}
\varphi_{1/\lambda}\sim -(\zeta_ca_p)f_c(1-r/a_p)
\label{e:limitwf}
\end{eqnarray}
outside the range of the strong interaction.

We  return to the neglect of the $K^0$--$K^\pm$ mass difference
$\Delta$ in calculating the logarithmic derivative $\zeta_n$
for the
$K^0\bar K^0$ channel. Including $\zeta_n =-\sqrt{2M_K\Delta}$ leads to a revised
version
of the eigenvalue equation  (\ref{e:eigenvalues}) that now reads
\begin{eqnarray}
1-\frac{1}{2}(\zeta_c+\zeta_n)\big(\frac{1}{\zeta_0}+\frac{1}{\zeta_1}\big)+\frac{\zeta_c\zeta_n}{\zeta_0\zeta_1}=0
\label{e:eigenrevised}
\end{eqnarray}
The lowest eigenvalue of this equation is $\lambda=0.9860+0.0061i$
instead of $0.9863+0.0082i$, an inessential change. The reason is the
dominating influence of the isoscalar channel for which
$\zeta_0 d\approx 18.0-10.5i$, that overshadows the remaining
logarithmic derivatives, which are all ${\cal O}(1)$, and  suppresses
the last term in Eq.~(\ref{e:eigenrevised}).
\subsection{Coulomb distortion}
It is also useful to document the modification due to the Coulomb 
interaction on the
low--momentum behavior of the absorption cross section as given by
the second of Eqs.~(\ref{e:sigmas}). Using Eqs.~(\ref{e:coulsmatrix}),  
(\ref{e:zetap}) and (\ref{e:zerolndrv}), one finds that
\begin{eqnarray}
 \sigma_a\approx\frac{4\pi}{k^2}Im(a^{eff}_p)Im(\zeta_c)
\approx - \frac{4\pi}{k}\frac{Im(a^{eff}_p)}{|f^{(-)}_c(k,d)|^2} 
\end{eqnarray}
since $Im(\zeta_c)=-k|f^{(-)}_c(k,d)|^{-2}$ 
is the imaginary part of the logarithmic derivative $\zeta_c$ of the 
incoming Coulomb wave. The latter relation is easily found by writing 
$f^{(-)}_c(k,r)={\rm exp}(i\sigma)(G(k,r)-iF(k,r))$
in terms of the regular and irregular Coulomb solutions 
 $F(k,r)\sim \sin(kr-\eta\ln2kr+\sigma)$, 
$G(k,r)\sim \cos(kr-\eta\ln2kr+\sigma)$, so that 
\begin{eqnarray}
\zeta_c=\Big(\frac{GG^\prime+FF^\prime-ik}{G^2+F^2}\Big)_{r=d}
\end{eqnarray} 
after using the
Wronskian relation $(GF^\prime-G^\prime F)=k$ that holds for all $r$.
 For $kd\sim \mu\alpha d << 1$ one finds from Eq.~(\ref{e:fc}) that
\begin{eqnarray}
|f_c^{(-)}(k,d)|^2 &&=
\Big(\frac{e^{2\pi\eta}-1}{2\pi\eta}\Big) 
\Big[\Big(1-2\mu\alpha d(\ln2\mu\alpha d+2\gamma-1)\Big)^2 
+
\Big(\frac{2\pi\mu\alpha d}{(e^{2\pi\eta}-1)}\Big)^2\Big]
\nonumber\\
&&\sim \Big(\frac{e^{2\pi\eta}-1}{2\pi\eta}\Big),
\quad \eta=-\mu\alpha/k
\label{e:coulcorrection}
\end{eqnarray}
in agreement with Ref.~\cite{LL58}.

\subsection{Detailed balance}
If one ignores  the real  one--boson exchange   
potential $V_I$ in Eq.~(\ref{e:newschr}), then the Jost function
for $K\bar K$ scattering due to the  annihilation part 
 alone is simply
$f(k)=\zeta+ik$. This gives rise to a $s$--wave $I=0$  scattering 
length of
 \begin{eqnarray}
4\pi a_0(K\bar K)=- M_Kc^2_0
\end{eqnarray}
after using the definition (\ref{e:zeta}) for $\zeta$.
 However, since  $\pi\pi$ annihilation into $K\bar K$  is described 
by the same diagram in Fig.~\ref{f:fig3} with the pion and kaon lines 
interchanged, the $I=0,\;\pi\pi$ 
 scattering length due to this process alone is also given to the same approximation by the 
  expression as above, but
 with $M_\pi$ and $M_K$ interchanged in the product $M_Kc^2_0$, i.e. 
\begin{eqnarray}
4\pi a_0(\pi\pi)=-[M_Kc^2_0]_{M_\pi\leftrightarrow M_K}=-\frac{k}{p}M_Kc^2_0
=4\pi a_0(K\bar K)\left( \frac{k}{p}\right)
\end{eqnarray}
We have used Eq.~(\ref{e:c0}) to perform the mass interchange. Here $p$ and $k$
are the relative momenta of a pion or kaon pair with the same total
CM energy $P_0$.
There is no contribution to the $a_2(\pi\pi)$ scattering length  
from Fig.~\ref{f:fig3} in
the $I=2$ channel, since the intermediate kaon loop can at most carry $I=1$.
Consequently detailed balance holds between the $I=0$ absorption 
cross section $\sigma^{(0)}_a(\pi\pi\to K\bar K)$ and its inverse 
$\sigma^{(0)}_a(K\bar K \to \pi\pi)$  when viewed in the 
isospin basis under the above approximations. 

However the physical $\pi^+\pi^-$ channel is described by 
the scattering length $a(\pi^+\pi^-)=\frac{2}{3} a_0(\pi\pi) 
+\frac{1}{3} a_2(\pi\pi)$. But 
$Im\big(a_2(\pi\pi)\big)=0$,
so
\begin{eqnarray}
\sigma_a(\pi^+\pi^-\to K\bar K)
=-\frac{4\pi}{p}Im\left( \frac{2}{3} a_0(\pi\pi)\right)
=\frac{2}{3}\frac{k^2}{p^2}\sigma^{(0)}_a(K\bar K\to \pi\pi)
\label{e:detail}
\end{eqnarray}
provided that $P_0\ge 2M_K$.

\subsection{Dimensional regularization of $J(P_0)$}
Replace the three--dimensional integral in Eq.~(\ref{e:gammahat1}) by its
analog in $d$ dimensions\cite{GLRG01},
\begin{eqnarray}
J(P_0)&&= \int\frac{d^3{\bf l}}{(2\pi)^3}\Big(\frac{1}{{\bf
l^2}/M+2M-P_0}\Big) \to\frac{1}{(4\pi)^{d/2}}\frac{1}{\Gamma(d/2)}\int_0^\infty
dL^2\frac{(L^2)^{d/2-1}}{L^2/M+2M-P_0}
\nonumber
\\
&&= \Big[M(2M-P_0)\Big]^{d/2-1}\frac{M}{(4\pi)^{d/2}}\Gamma[1-d/2]
\label{e:d}
\end{eqnarray}
or, as $d\to 3$,
\begin{eqnarray}
J(P_0)=-\frac{M}{4\pi}\Big[M(2M-P_0\Big]^{1/2}=
\frac{M}{4\pi}\Big[M|(P_0-2M)|\Big]^{1/2} e^{i(\phi+\pi)/2}
\quad 0<\phi<2\pi
\label{e:JP}\quad JP
\end{eqnarray}
on the first sheet of the complex $P_0$ plane.  Note that $J(P_0)$ is pure imaginary and
positive on the upper lip of the cut $\phi=0$  extending from $2M$ to $\infty$.
The analytic continuation to the second sheet through the
cut starting at the branch point $P_0=2M$ is obtained by
taking $-2\pi<\phi<0$.

\begin{figure}
\resizebox{0.88\textwidth}{!}{%
\rotatebox{-90}{
 \includegraphics{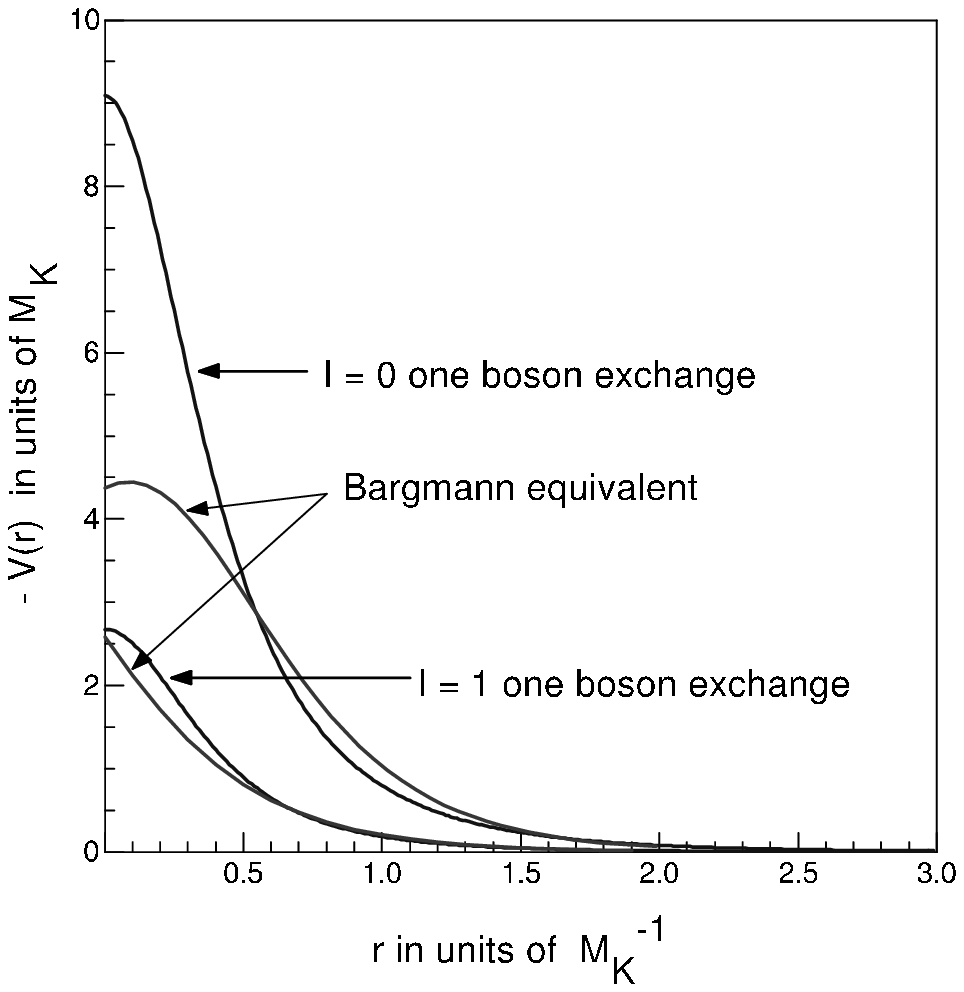}
}}
\caption{\label{f:fig1}
One boson exchange potentials using a cutoff of $\Lambda$ =
4 GeV. The coupling constant is $\alpha_s=2.9$ and the masses of the exchanged
mesons are ($M_\rho,M_\omega,M_\phi)=(769,783,1019)$ MeV. The analytical
Bargmann potentials that reproduce the same scattering lengths and effective ranges
 as the original one boson exchange potentials are shown as companion curves.}
\end{figure}

\begin{figure}
\resizebox{0.88\textwidth}{!}{%
\rotatebox{-90}{
 \includegraphics{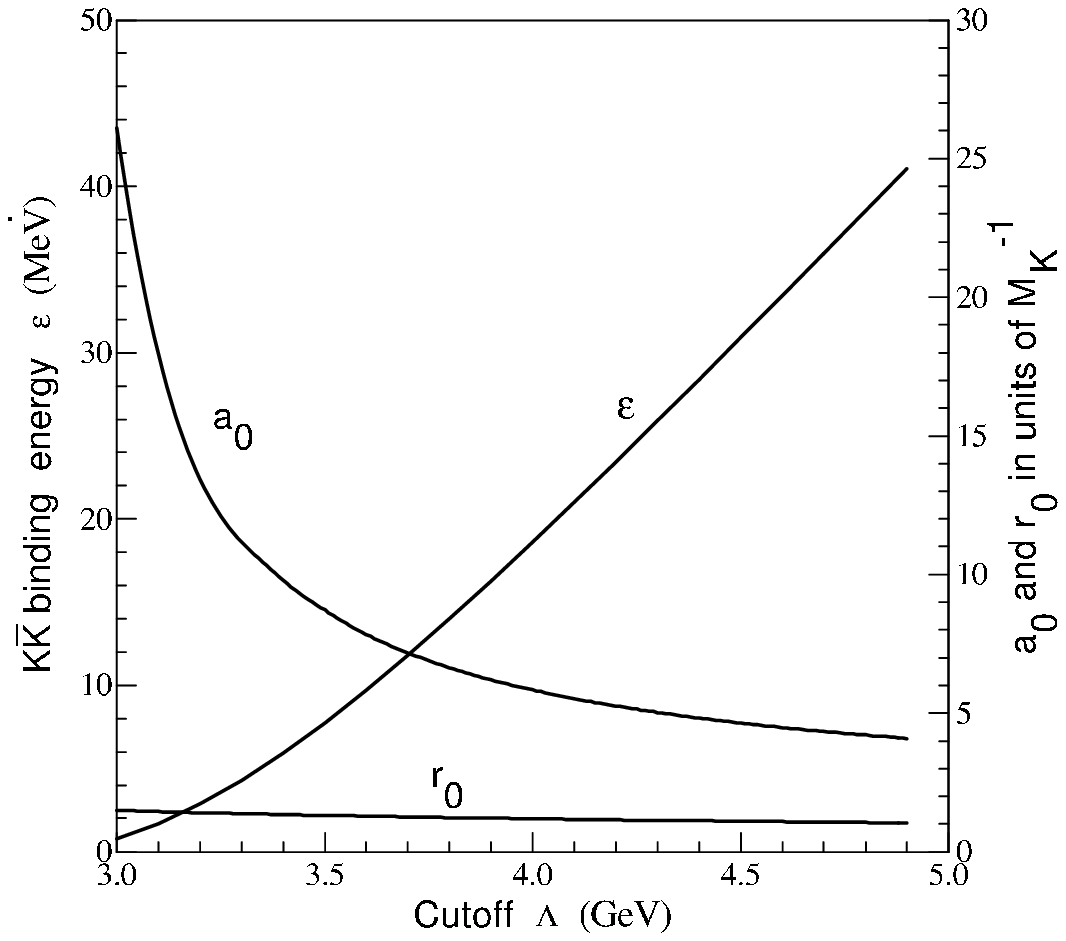}
}}
\caption{\label{f:fig2}
Dependence of the $K\bar K$ binding energy $\varepsilon$
(left hand scale), scattering length $a_0$ and effective range $r_0$
(right hand scale) on the choice of cutoff $\Lambda$ used in the form factor
for the isosalar one boson exchange potential.}
\end{figure}

\begin{figure}
\resizebox{0.88\textwidth}{!}{%
\rotatebox{-90}{
 \includegraphics{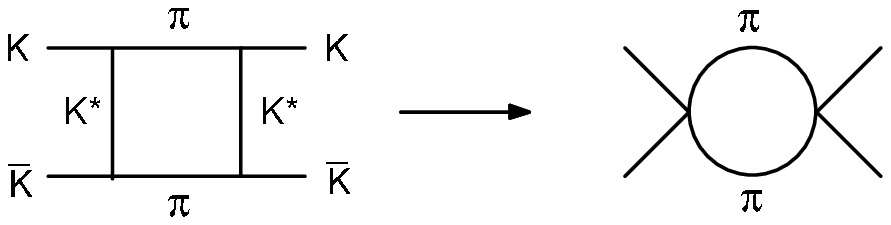}
}}
\caption{\label{f:fig3}
 Replacement of the irreducible $K\bar K$
scattering diagram involving the exchange of two $K^*$'s by
 an equivalent  pion loop diagram  with point vertices.}
\end{figure}

\begin{figure}
\resizebox{0.88\textwidth}{!}{%
\rotatebox{-90}{
 \includegraphics{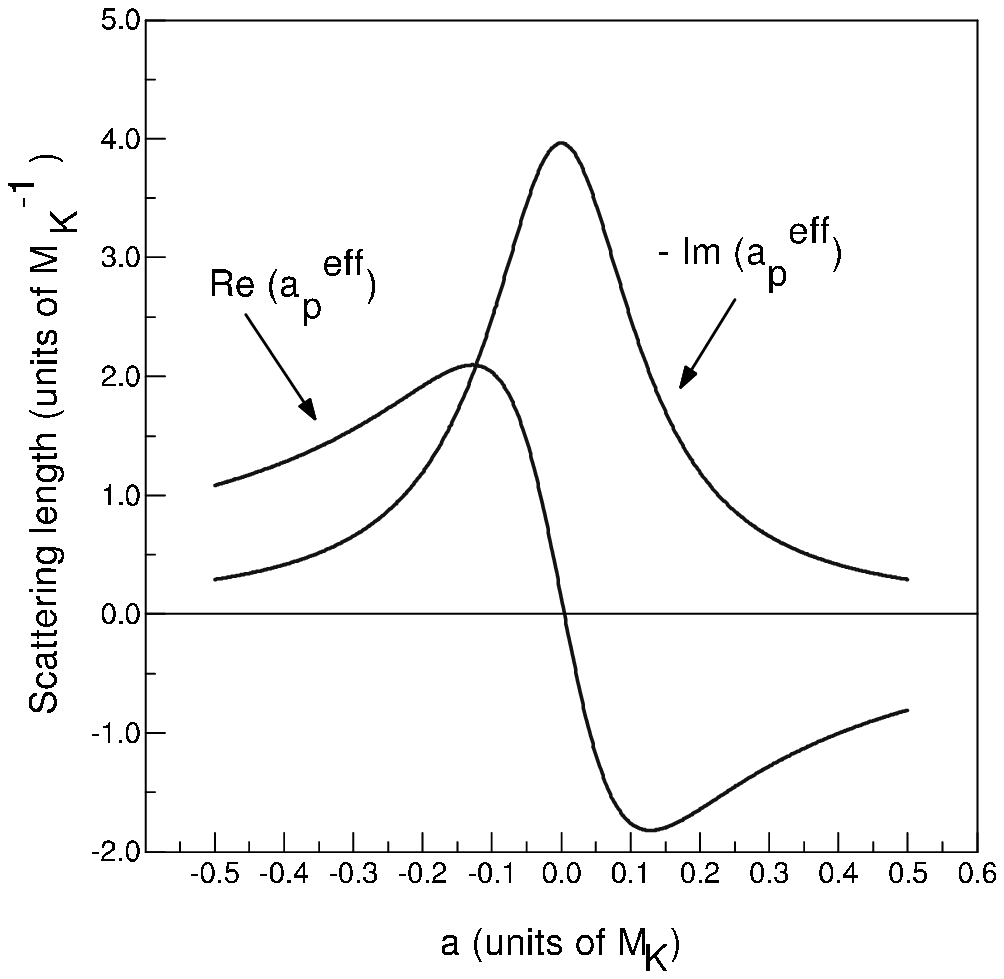}
}}
\caption{\label{f:fig4}
Real and imaginary parts of the complex scattering length $a^{eff}_p$
as a function of the parameter $a$ of the isoscalar Bargmann potential.
The remaining parameters are as in Table~\protect\ref{t:table1}.
The  $f_0(980)$ ground state becomes 
unbound as Re$(a^{eff}_p)$ passes through zero from positive 
to negative values.}
\end{figure}

\begin{figure}
\resizebox{0.88\textwidth}{!}{%
\rotatebox{-90}{
 \includegraphics{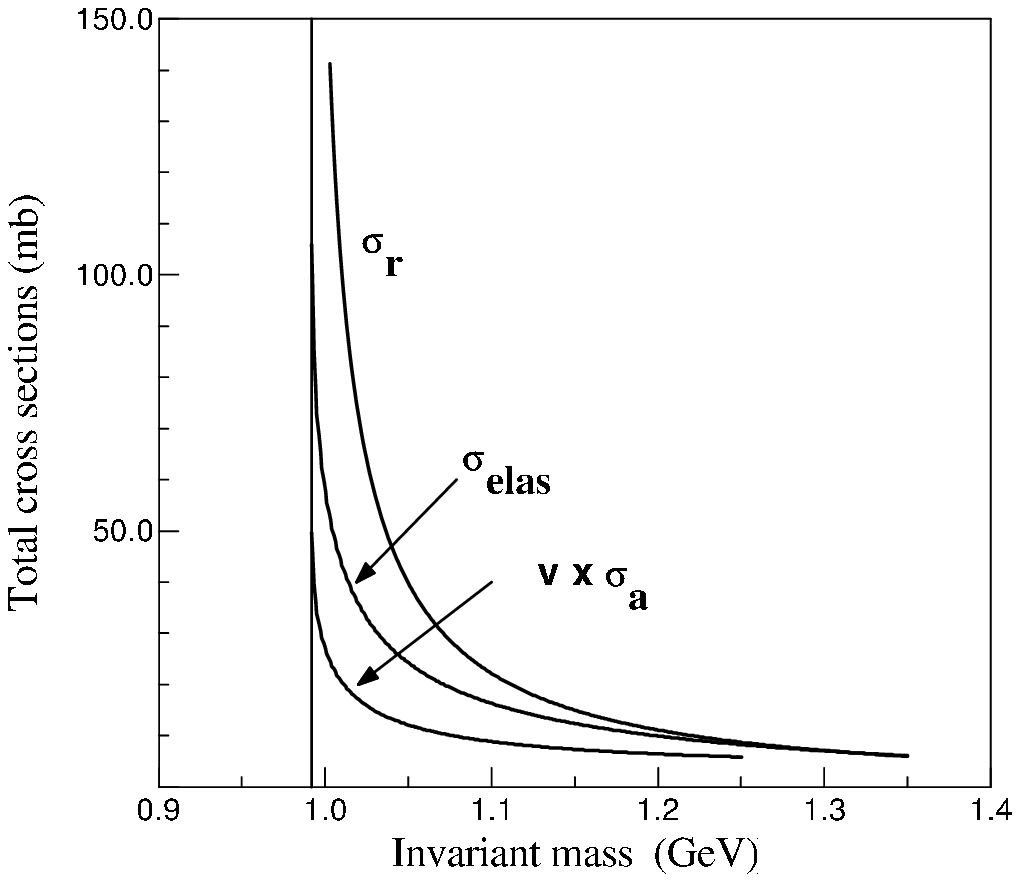}
}}
\caption{\label{f:fig5}
Calculated total $s$--wave elastic, reaction and $v\times$ the    
annihilation  cross sections  
 as a function of the invariant mass $P_0$  
of the colliding kaon--antikaon pair. The vertical line indicates the 
threshold value at $2M_K$.} 
\end{figure}

\begin{figure}
\resizebox{0.88\textwidth}{!}{%
\rotatebox{0}{
 \includegraphics{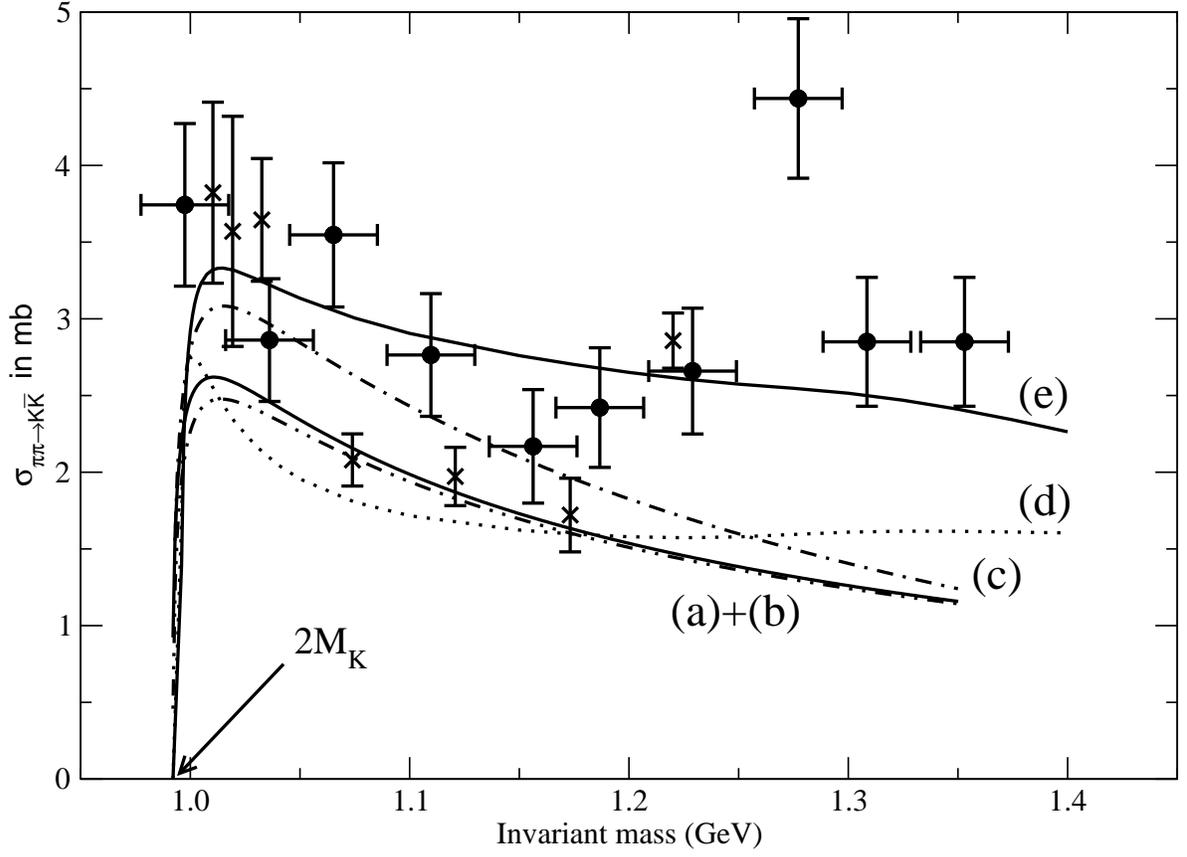}
}}
\caption{\label{f:fig6}
The $\protect\pi^+\pi^-\to K\bar K$ cross section given by detailed 
balance, Eq.~(\protect\ref{e:detbalance}),  as a function of the 
$\protect K\bar K$ invariant mass 
$P_0$. The two curves designated (a)+(b) give the results
with (upper curve) and without (lower curve) Coulomb distortion. 
Also shown are the $\protect\pi^+\pi^-\to K\bar K$ 
experimental data\protect\cite{SDP73} (closed circles), and 
\protect\cite{ADM77} (crosses). 
Curve (c)) is a fit to these data
using the parametrized  scattering matrix of
\protect\cite{SDP73}. Curves $(d)$ and $(e)$ correspond to coupled
channel calculations of the $\pi^+\pi^-\rightarrow K\bar{K}$ cross
section with (curve $(e)$) and without (curve $(d)$) $f_0^\prime$ mixing.} 
\end{figure}

\begin{figure}
\resizebox{0.88\textwidth}{!}{%
\rotatebox{0}{
 \includegraphics{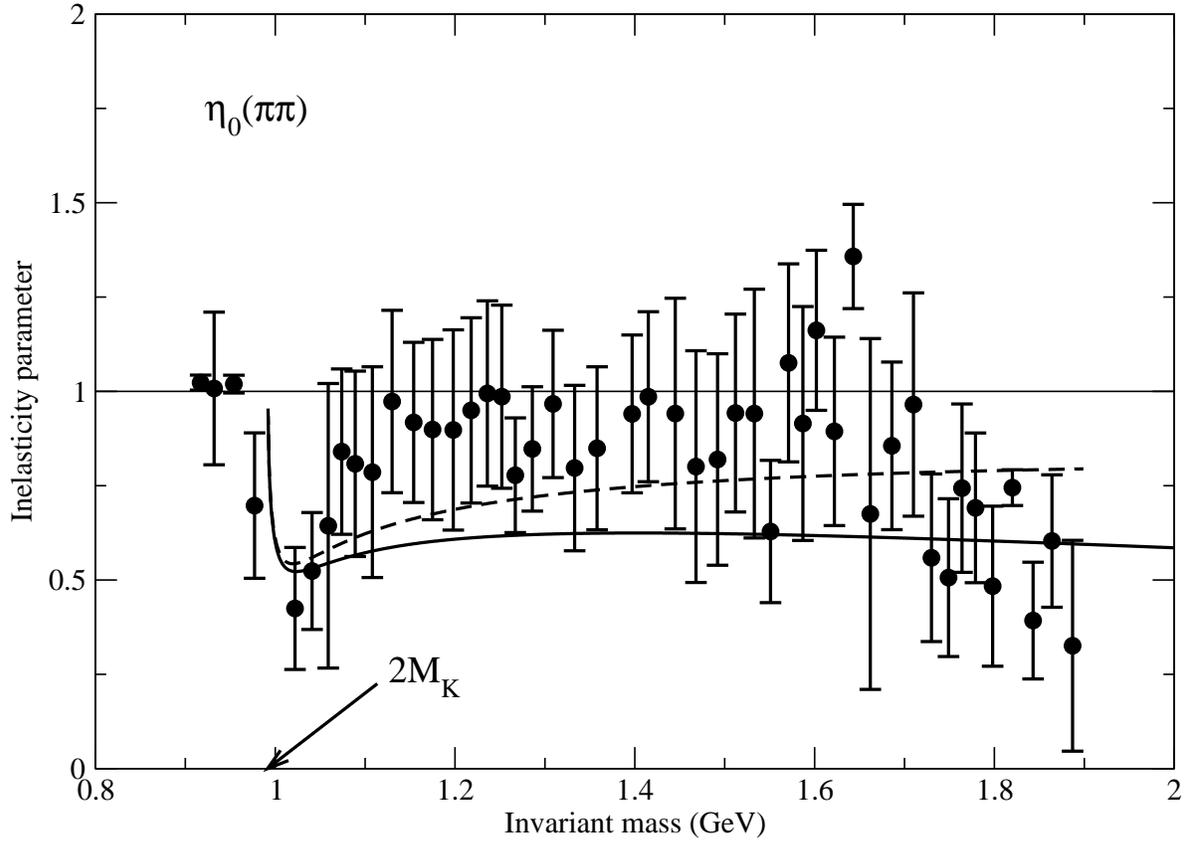}
}}
\caption{\label{f:fig6a}
The $\pi\pi$ inelasticity parameter $\eta_0(\pi\pi)$ as a function of the
CM energy. The experimental data is taken from \cite{BHyams1973}. 
The upper curve
shows the behavior of the inelasticity given by a fully relativistic
calculation of the pion loop integral.
} 
\end{figure}

\begin{figure}
\resizebox{0.88\textwidth}{!}{%
\rotatebox{-90}{
 \includegraphics{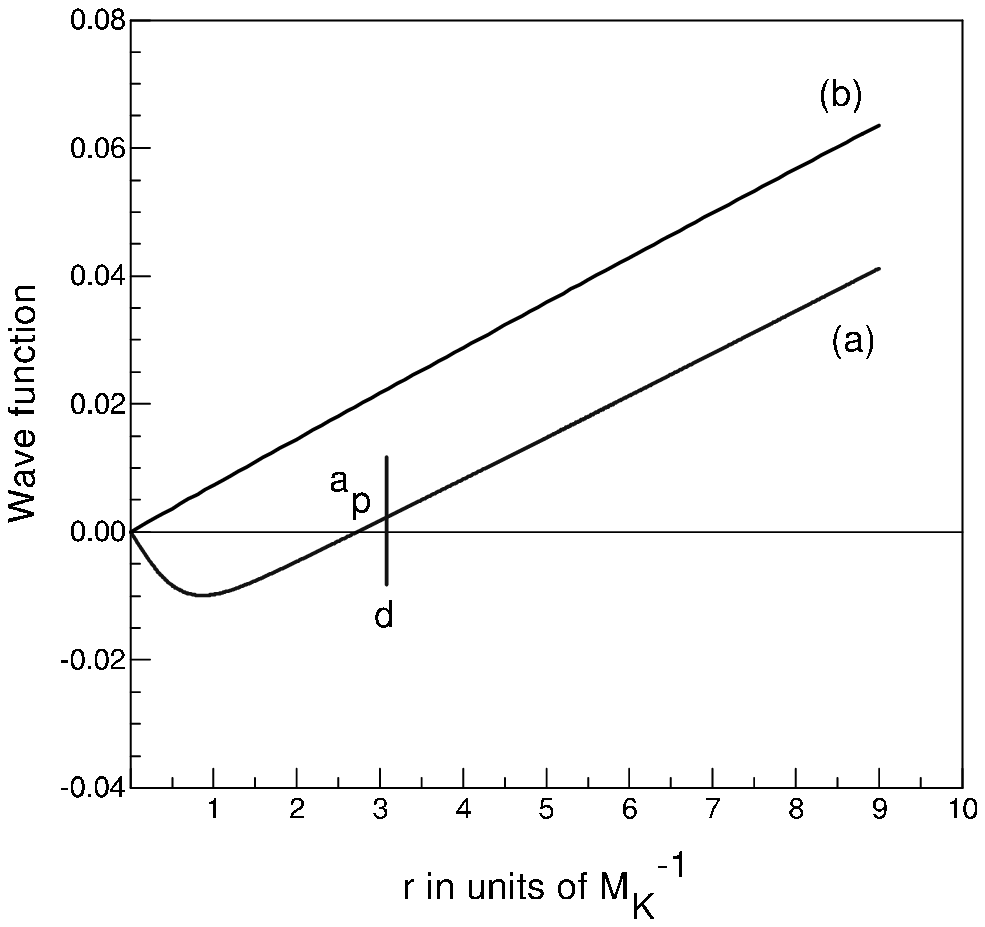}
}}
\caption{\label{f:fig7}
Kaonium ground state radial wave functions: (a) with and (b) without
the  meson exchange fields present. In (a) the strong field solution,
Eq.~(\protect\ref{e:kaoniumwf}), is joined
smoothly at $r=d$ onto  the Whittaker function $W_{1/\lambda,\frac{1}{2}}
(2\lambda\mu\alpha r)$ for $\lambda=0.981$. The node in this solution falls at the $K^+K^-$ scattering
length $a_p$. In (b) the solution
$W_{1,\frac{1}{2}}(2\mu\alpha r)
=2\mu\alpha r\: \exp{(-\mu\alpha r)} $ for a pure Coulomb field, $\lambda=1$,
is plotted for the same range of $r$. To set the scale, note
  that the Bohr radius of kaonium
would fall at $\sim 300$ units to the right of this  figure.}
\end{figure}

\begin{figure}
\resizebox{0.88\textwidth}{!}{%
\rotatebox{-90}{
 \includegraphics{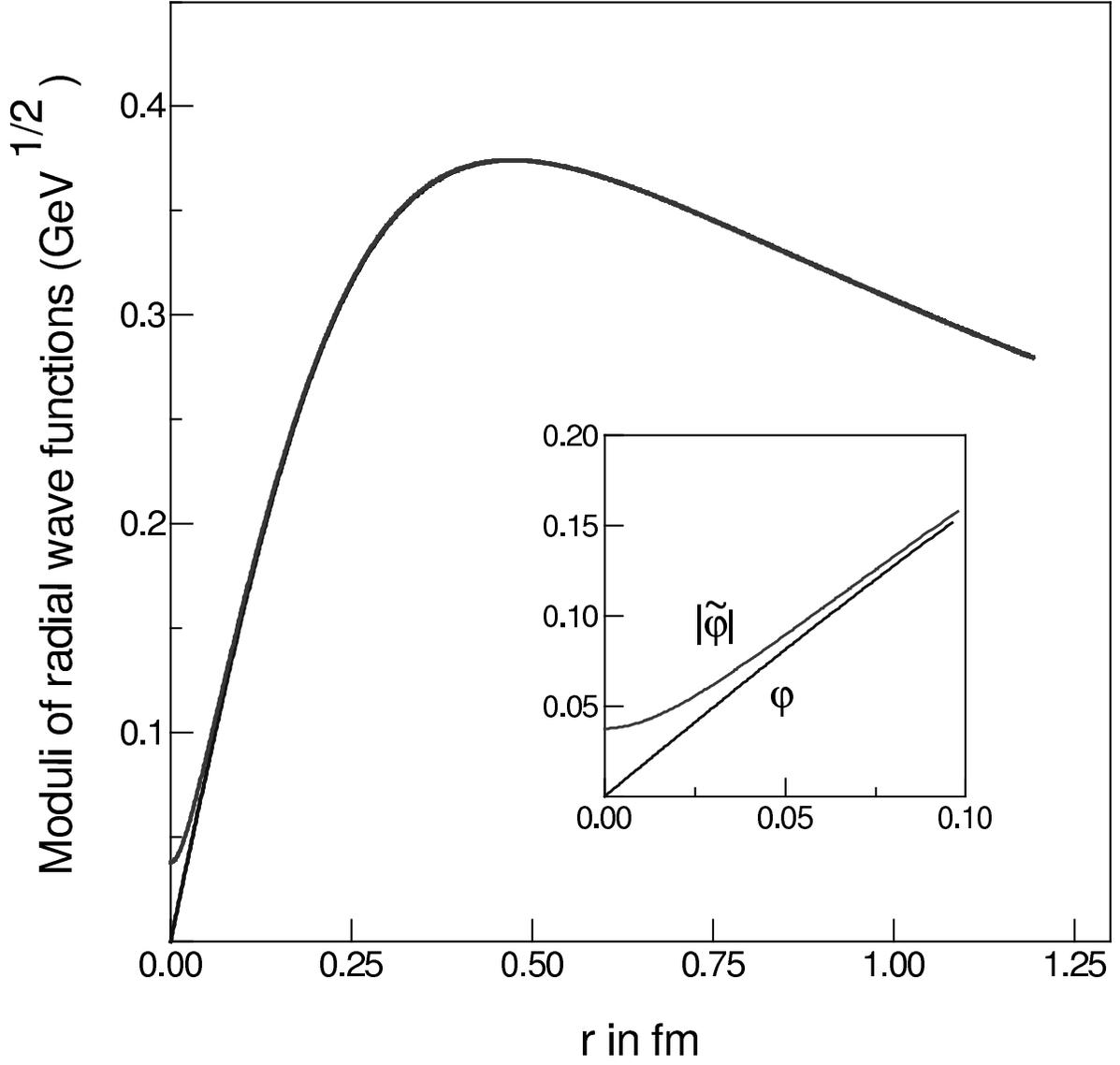}
}}
\caption{\label{f:fig7a}
A plot of (the moduli of) the normalized radial wave functions 
$\tilde\varphi(k_r,r)$ and 
$\varphi(-i\kappa,r)$ that include and exclude annihilation effects respectively. The inset
details their behavior for small $r$. Notice that 
$|\tilde\varphi(k_r,r)|$ ``heals'' to $\varphi(-i\kappa,r)$ 
within $ 0.1 $ fm away from the origin.
}

\end{figure}
\begin{figure}
\resizebox{0.88\textwidth}{!}{%
\rotatebox{-90}{
 \includegraphics{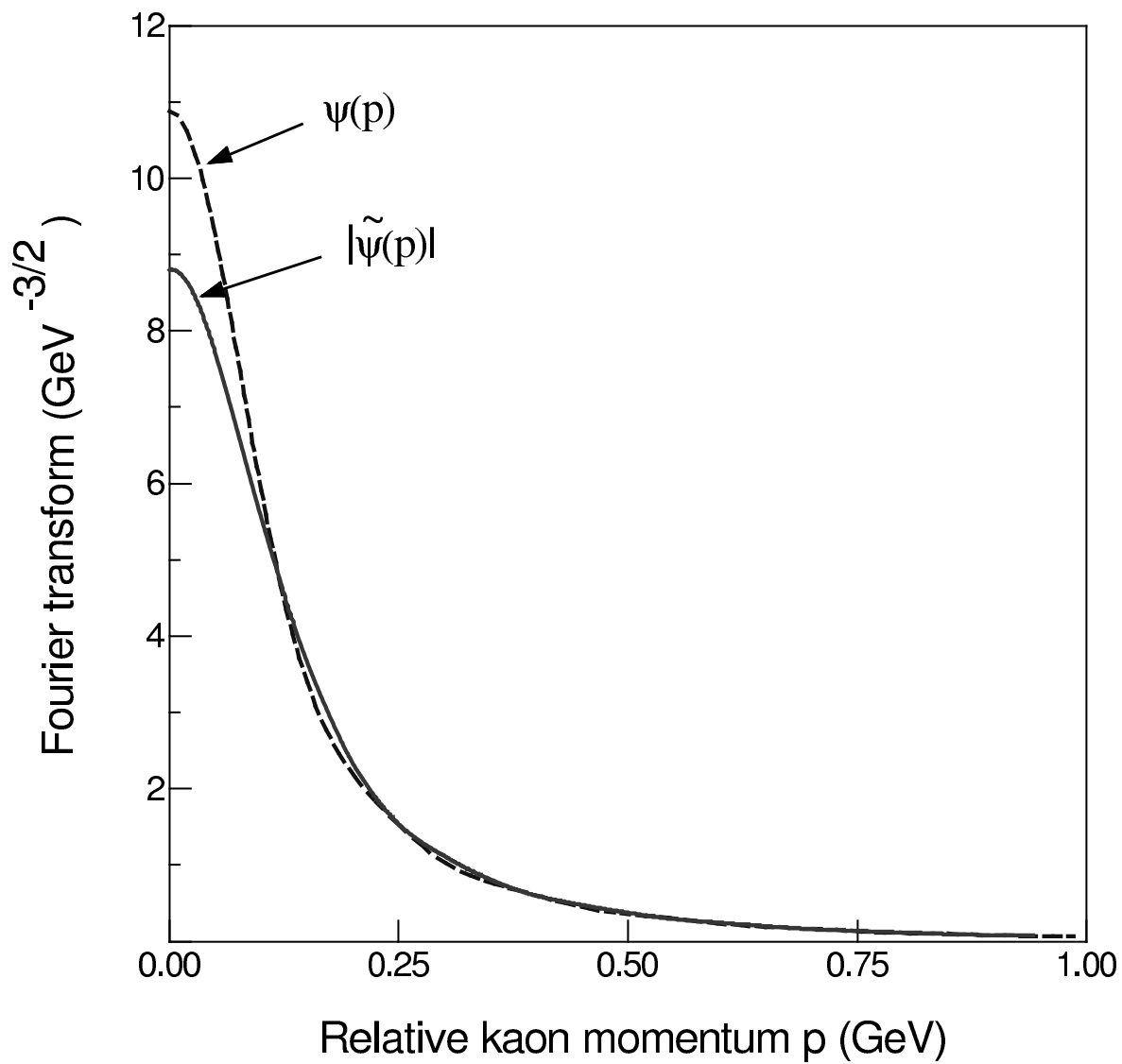}
}}
\caption{\label{f:fig7b}
 Normalized fourier transforms of the spatial wave functions $\tilde\psi(r)$
 with and $\psi(r)$ without annihilation.
}
\end{figure}

\begin{figure}
\resizebox{0.88\textwidth}{!}{%
\rotatebox{0}{
 \includegraphics{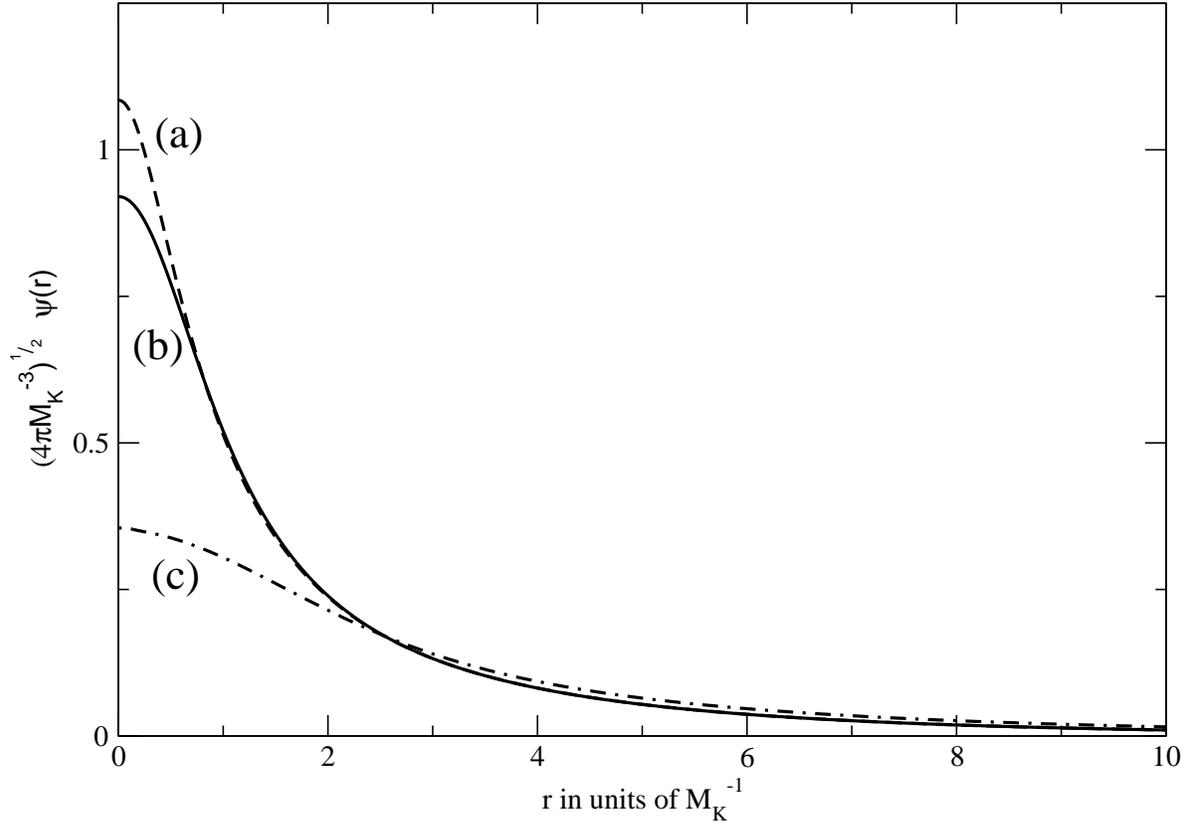}
}}
\caption{\label{f:fig8}
Comparison of the bound state wave function obtained numerically
(curve $(a)$)
for the isoscalar one boson exchange potential of Eq.~(\ref{e:OBE}),
with that
given by the analytical solution, Eq.~(\ref{e:bargu}), for the equivalent
Bargmann potential (curve $(b)$). Curve $(c)$ shows the wave function in a gaussian
potential adjusted to give a binding energy of $10\textrm{ MeV}$ \cite{Barnes1985}.}
\end{figure}

\begin{figure}
\resizebox{0.88\textwidth}{!}{%
\rotatebox{0}{
 \includegraphics{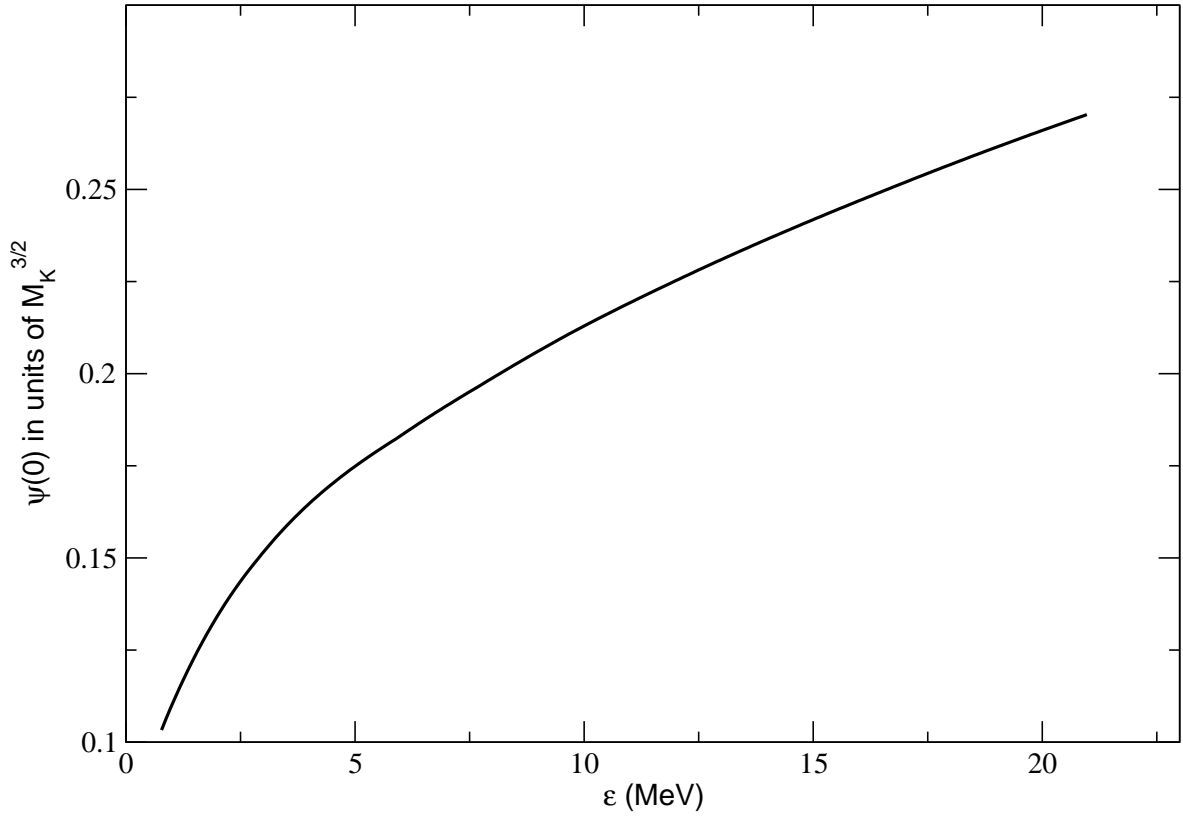}
}}
\caption{\label{f:fig9} Variation of the spatial wave function
  $\psi(0)$, Eq.~(\ref{e:bargu}), with binding energy.
}
\end{figure}

\begin{center}
\begin{table}
\caption[]{Parameters of the Bargmann potentials,
Eq.~(\ref{e:bargpot}), that give rise to identical  scattering lengths and effective ranges
as the one boson exchange potentials in Eqs.~(\ref{e:OBE})
and (\ref{e:OBET}).
The last column gives the value of the logarithmic
derivative that determines  the strength
the annihilation potential. The units are $M_K$.}
\label{t:table1}
\begin{ruledtabular} \begin{tabular}{cccc}
Isospin&$a$ &  $b$ & $\zeta$\\
\colrule
$I=0$ & $-0.1936$ & $1.491$  & $-17.409i$\\
$I=1$ & $+0.9219$ & $1.462$  & $-26.114i$\\
\end{tabular} \end{ruledtabular}
\end{table}
\end{center}

\begin{center}
\begin{table}
\caption[]{Summary of calculated isoscalar $K\bar K$ scattering lengths
$a^\prime_0$ and effective ranges $r^\prime_0$ (in units $M^{-1}_K$), based on the $a$ and $b$ Bargmann potential
parameters (in units $M_K$) that reproduce the
 complex bound state  at $P_0$ shown in column two.
 The physical constants that enter into these calculations
 are: $(M_\pi,M_K,M_{K^*})=(140,496,892)$ MeV, and
 $\alpha_s=g^2_{\rho\pi\pi}/4\pi\approx 2.9$,
 using the KSRF value of $g_{\rho\pi\pi}\approx 6$. The last four
 rows  list the values of
  $a$ and $b$, and the resulting isoscalar scattering lengths and effective
  ranges 
  that would be required to reproduce  the theoretical
  results of Weinstein and Isgur (WI), the fits of
   Morgan and Pennington (MP), and the experimental data.}

   \label{t:table2}

\begin{ruledtabular} \begin{tabular}{cccccc}
Source&$P_0=M-i\Gamma/2$ (MeV) &  $a$  & $b$ & $a^\prime_0$ & $r^\prime_0 $\\
\colrule
This calculation& $981-25i$ & $-0.194$ &  $1.491$ & $4.28$ & $3.20$\\
WI\cite{WS90}&$972-16i$ & $-0.213$ &  $1.170$ & $5.01$ & $2.12$\\
MP\cite{MP93}&$970-24i$ & $-0.231$ &  $1.367$ & $4.30$ & $2.16$\\
PDG\cite{PDG92}&$(974\pm 3)-(24\pm 5)i$ & $-0.22\pm 0.02$ &  $1.41\pm 0.16$
& $4.35\pm 0.49$ & $2.46\pm 0.35$\\
 E791 \cite{fermilab2001}&$(975\pm 3)-(22\pm 2)i$&$-0.21\pm0.02$&$1.37\pm 0.05$&
$4.52\pm 0.20$&$2.51\pm 0.33$
\end{tabular} \end{ruledtabular}
\end{table}
\end{center}

\begin{center}
\begin{table}
\caption[]{Complex scattering lengths and effective ranges for
isospin zero and one scattering channels as calculated from 
Eq.~(\ref{e:cmplxln}), as well as from coupled channel calculations
with the $f_0^\prime$ and the $\pi \eta$-channel\cite{julich03} included (in units of $M^{-1}_K$).}
\label{t:table3}
\begin{ruledtabular} \begin{tabular}{cccc}
 &Isospin&$a^{eff}_I$&$r^{eff}_I$\\
\colrule
One boson exchange
  &$\begin{array}{c} I=0 \\ I=1 \end{array}$
  &$\begin{array}{c} 4.281-2.398i \\ -0.401 \end{array}$
  &$\begin{array}{c} 1.169-0.178i \\ 3.702 \end{array}$ 
\\ \hline
$+f_0^\prime$ mixing $+\pi\eta$-channel
  &$\begin{array}{c} I=0 \\ I=1 \end{array}$
  &$\begin{array}{c} $2.165-4.592i$ \\ $0.147-1.944i$ \end{array}$ 
  &$\begin{array}{c} $0.151-0.404i$ \\ $-1.463-0.202i$ \end{array}$ \\

\end{tabular} \end{ruledtabular}                
\end{table}
\end{center}

\begin{center}
\begin{table}
\caption[]{ Complex level shifts and lifetimes due to the strong interactions away from the point Coulomb 
value $-\frac{1}{4}\alpha^2M_K/n^2=-6.607/n^2$ keV
for the ground and excited $s$--states of kaonium. Numerical 
solutions of the Kudryavtsev--Popov equation have been generated for
   the $K^+K^-$
   scattering length  $a^{eff}_p=(1.940-1.199i)M^{-1}_K$ given 
   by Eq.~(\ref{e:apeff}). 
   The  matching
   radius is $d\sim 3M^{-1}_K$. The real part of $\lambda$ lies
   close the sequence $1/1,1/2,1/3,1/4,\cdots$ for a pure Coulomb spectrum.}
\label{t:table4}
\begin{ruledtabular} \begin{tabular}{cccc}
Level&$\lambda$&$\Delta E-i\Gamma/2$ (keV)&Lifetime ($\times 10^{-18}$ sec)\\
\colrule
{$3^{rd}$}&$0.2491+0.0005i$&$0.003-0.002i$& $199$ \\ 
{$2^{nd}$}&$0.3318+0.0009i$&$0.007-0.004i$&$84$ \\
{$1^{st}$}&$0.4965+0.0020i$&$0.023-0.013i$&$25$ \\
{Ground}&$0.9863+0.0079i$&$0.180-0.103i$&$3.2$ \\ 
\end{tabular} \end{ruledtabular}                
\end{table}
\end{center}

\begin{center}
\begin{table}
\caption[]{A repeat of the calculations described in Table~\ref{t:table4}, but for 
   the scattering length  $\hat{a}^{eff}_p=(1.156-3.268i)M^{-1}_K$ of 
   Eq.~(\ref{e:felixapeff}) which has been taken from the model of  
Ref.\cite{julich03} that includes 
   $f_0^\prime$ mixing and the $+\pi\eta$-channel.
 This choice leads to lifetimes of about a factor three
   shorter than those given in Table~\ref{t:table4}.}
\label{t:table5}
\begin{ruledtabular} \begin{tabular}{cccc}
Level&$\lambda$&$\Delta E-i\Gamma/2$ (keV)&Lifetime ($\times 10^{-18}$ sec)\\
\colrule
{$3^{rd}$}&$0.2494+0.0014i$&$0.002-0.005i$&$71$ \\
{$2^{nd}$}&$0.3322+0.0025i$&$0.005-0.011i$&$30$ \\
{$1^{st}$}&$0.4975+0.0056i$&$0.017-0.037i$&$8.9$\\
{Ground}&$0.9899+0.0223i$&$0.137-0.291i$&$1.1$\\ 
\end{tabular} \end{ruledtabular}                
\end{table}
\end{center}

\begin{center}
\begin{table}
\caption[]{A comparison of predicted lifetimes of the kaonium ground state
for increasing complexity of the $K^+K^-$ interaction.}
\label{t:table6}
\begin{ruledtabular} \begin{tabular}{ccc}
& Scattering length (fm) & $\tau (K^+K^-\to \pi\pi)$  (sec)\\
\colrule
Free annihilation at rest& $0.00-0.01i$ & $1.2\times 10^{-16}$ \\
+One boson exchange  & $0.77-0.48i$ & $3.2\times 10^{-18}$\\
+ $f_0^\prime$ mixing $+\pi\eta$-channel\cite{julich03}& $0.46-1.30i$ & $1.1\times 10^{-18}$\\
\end{tabular} \end{ruledtabular}
\end{table}
\end{center}


\newpage
\begin{thebibliography}{9}


\bibitem{amsler}
C. Amsler, Rev. Mod. Phys. {\bf 70}, 1293 (1998).

\bibitem{godfrey}
S. Godfrey and J. Napolitano, Rev. Mod. Phys. {\bf 71}, 1411 (1999).


\bibitem{Montanet2000}
L. Montanet, Nucl.Phys. {\bf B86}, 381 (2000).

\bibitem{Anisovich2000}
V.V. Anisovich et al., Phys. Lett. {\bf B480}, 19 (2000).

\bibitem{achasov82}
N.N. Achasov and V. N. Ivanchenko, Nucl. Phys. {\bf B315}, 465 (1989).


\bibitem{WS90}
J. Weinstein and N. Isgur, Phys. Rev. Lett. {\bf 48}, 659 (1982);
Phys. Rev. {\bf D 27}, 588 (1983); Phys. Rev. {\bf D 41}, 2236 (1990).



\bibitem{achassov00}
N.N. Achasov et al.,  Phys. Lett. {\bf B485}, 349 (2000).

\bibitem{KLOE} 
A. Aloisio et al.,  Phys. Lett. {\bf B537}, 21 (2002).

\bibitem{oller03}
J. A. Oller, Nucl. Phys.{\bf A714}, 161 (2003).

\bibitem{achasov03}
N.N. Achasov, hep-ph/0309118


\bibitem{gams}
D. Alde et al. (GAMS), Z. Phys.  {\bf C66}, 375 (1995).

\bibitem{gunter01}
J. Gunter et al. (E852), Phys. Rev. {\bf D64}, 072003 (2001).


\bibitem{Achasov1}
N.N. Achasov et al., Phys. Rev. {\bf D58}, 054011 (1998).


\bibitem{sassen03}
F. P. Sassen et al., Phys. Rev.  {\bf D68}, 036003 (2003).

\bibitem{LLR02}
E. Lipartia, V. E. Lyubovitskij, A. Rusetsky, Physics Lett. {\bf B 533},
285 (2002).

\bibitem{GLRG01}
J. Gasser, V. E. Lyubovitskij, A. Rusetsky, and A. Gall, Phys. Rev. {\bf D 64},
016008 (2001).

\bibitem{HadAtom02}
For a recent review, see ''Proceedings of the International Workshop on 
Hadronic Atoms'', CERN, Geneva,
October 14--15, 2002, ed. L. Afanasyev, A. Lanaro, J. Schacher,
arXiv:hep--ph/0301266 v1.


\bibitem{CMY99}
A. Czarnecki, K. Melnikov, A. Yelkhovsky, Phys. Rev. {\bf A 59}, 4316 (1999).

 \bibitem{deSwart63}
 J.J. de Swart, Rev. Mod. Phys.{\bf 35}, 916 (1963).


\bibitem{Lohse90}
D. Lohse, J. W. Dorso, K. Holinde, J. Speth, Nucl. Phys., {\bf A516}, 513 (1990).

\bibitem{Jansen95}
G. Jan\ss en, B. D. Pearce, K. Holinde, J. Speth, Phys. Rev. {\bf D 52},
2690 (1995).

\bibitem{KSRF66}
 K.Kawarabayashi, M. Susuki, Phys. Rev. Lett. {\bf 16}, 255 (1966);
 X. Riazuddin, X. Fayyazuddin, Phys. Rev. {\bf 147}, 1071 (1966).

\bibitem{Sassen01}
F. P. Sassen, Diplomarbeit,
Rheinischen Freidrich--Wilhelms--Universit\"at, Bonn (2001), (unpublished).


\bibitem{Bargmann49}
V. Bargmann, Rev. Mod. Phys.{\bf 21}, 488 (1949).



\bibitem{Newton56}
R. G. Newton, Phys. Rev. {\bf 105}, 763 (1956).

\bibitem{fermilab2001}
 E. M. Aitala {\it et al.} (Fermilab E791 Collab.), Phys. Rev. Lett. {\bf 86},
 765 (2001).

\bibitem{SDP73}
S. D. Protopopescu {\it et al.}, Phys. Rev. {\bf D 7}, 1279 (1973).

\bibitem{ADM77}
A. D. Martin {\it et. al.}, Nucl. Phys. {\bf B 121}, 514 (1977)

\bibitem{BHyams1973}
B. Hyams {\it et al.} Nucl. Phys. {\bf B 64}, 134 (1973)


\bibitem{Deser54}
S. Deser, M. L. Goldberger, K. Baumann and W. Thirring, Phys. Rev. {\bf 96},
774 (1954).


\bibitem{KP79}
A. E. Kudryavtsev and V. S. Popov, JEPT Lett. {\bf 29}, 280 (1979);
V.S. Popov, A. E. Kudryavtsev and V.D. Mur, Sov.Phys. JETP {\bf 50},
865 (1979).



\bibitem{LL82}
See, for example, L. D. Landau and E. M. Lifshitz, {\it Quantum Electrodynamics}, Course
of Theoretical  Physics, Vol. 4, Second Edition, (Pergammon Press, 1982 ).




\bibitem{Newton60}
R. G. Newton, J. Math. Phys. {\bf1}, 319 (1960).

 \bibitem{Petersen71}
 J. L. Peterson, Physics Reports {\bf 2 C}, 151 (1971).


\bibitem{PDG92}
Particle Data Group, K. Hikasa {\it et al.}, Phys. Rev. {\bf D 45}, S1 (1992).


 \bibitem{MP93}
 D. Morgan and M. R. Pennington, Phys. Lett., {\bf B 258}, 444 (1991);
 Phys. Rev. {\bf D 48}, 1185 (1993).
M.Boglione and M. R. Pennington, Phys. Rev. {\bf D65}, 114010 (2002).


\bibitem{RHLRT2002}

R. H. Lemmer and R. Tegen, Phys. Rev. {\bf C 66}, 065202 (2002).

\bibitem{IZ80}
C. Itzykson and J. Zuber, {\it Quantum Field Theory} 
(McGraw--Hill, New York, 1980).

\bibitem{Barnes1985}
T. Barnes, Phys. Lett. {\bf 165B}, 434 (1985).

\bibitem{PRD96}
K. Hagiwara et al. (Particle Data Group), Phys.Rev. {\bf D66}, 010001 (2002).

\bibitem{Achasov82}
N.N. Achasov, S.A. Devyanin, and G.N. Shestakov, Phys. Lett. {\bf
  108B} 134, (1982); 
N.N. Achasov, Phys. Usp. {\bf 41}, 1149 (1998).

\bibitem{DR65}
D. Robson, Phys. Rev. {\bf 137}, 535 (1965).

\bibitem{RHL69}
R. H. Lemmer, {\it Cargese Lectures in
Physics}, edited by M. Jean (Gordon and Breach, 1969), Vol. 3, p. 483.



\bibitem{SWycech}
S. Wycech, and A.M. Green, Nucl. Phys. {\bf A562}, 446 (1993);
S.V. Bashinsky, and B.O. Kerbikov, Phys. At. Nucl. {\bf 59}, 1979 (1996).

\bibitem{julich03}
O. Krehl, R. Rapp, and J. Speth, Phys. Lett. {\bf B390}, 23 (1997).

\bibitem{RGN66}
See, for example, R. G. Newton, {\it Scattering Theory of Waves and Particles},
(McGraw--Hill Book Company, New York, 1966).

\bibitem{LL58}
L.D. Landau and E. M. Lifshitz, {\it Quantum Mechanics}, Vol. 3, (Pergamon Press
Ltd., London. 1958).

\bibitem{GR80}
See, for example, I. S. Gradshteyn and I. M. Ryzhik,
{\it Table of Integrals, Series, and Products},
(Academic Press, New York, 1980).


\end{thebibliography}
\end{document}